\documentclass[a4paper,11pt]{article}
\pdfoutput=1 

\usepackage{jheppub}

\usepackage[T1]{fontenc}
\usepackage[svgnames]{xcolor}
\usepackage{parskip}
\usepackage{float}
\usepackage{breqn}
\usepackage{comment}
\usepackage{float}
\usepackage{subcaption}
\allowdisplaybreaks
\usepackage{colortbl}

\usepackage{slashed}
\usepackage{bm}
\graphicspath{{plots/}{JCAP/plots/}}

\newcommand{\be}{\begin{equation}\begin{aligned}}
\newcommand{\ee}{\end{aligned}\end{equation}}
\newcommand{\nn}{\nonumber}

\renewcommand{\l}{\langle}
\renewcommand{\r}{\rangle}

\def\bea{\begin{eqnarray}}
\def\eea{\end{eqnarray}}
\def\nn{\nonumber}

\def\ie{{\it i.e.~}}
\newcommand{\vol}{\mathcal{V}}

\title{\boldmath Effect of Moduli Redefinitions on Fibre Inflation}

\author[a]{Dibya Chakraborty}
\author[b]{, Mishaal Hai}
\author[c]{, Sayeda Tashnuba Jahan}
\author[d,c]{, Ahmed Rakin Kamal}
\author[e]{, Md Shaikot Jahan Shuvo}

\affiliation[a]{Centre for Strings, Gravitation and Cosmology, Department of Physics,
Indian Institute of Technology Madras, Chennai 600036, India}
\affiliation[b]{\small Department of Computer Science and Engineering,
BRAC University, Kha 224, Bir Uttam Rafiqul Islam Ave, Dhaka 1212, Bangladesh}
\affiliation[c]{\small Department of Mathematics and Natural Sciences,
BRAC University, Kha 224, Bir Uttam Rafiqul Islam Ave, Dhaka 1212, Bangladesh}
\affiliation[d]{Department of Theoretical Physics and Astrophysics, Faculty of Science,
Masaryk University, Kotl\'a\v{r}sk\'a 2, CS-61137 Brno, Czechia}
\affiliation[e]{\small Initiative for the Theoretical Sciences, The Graduate Center, CUNY,
365 Fifth Ave, New York, NY 10016, USA}
\emailAdd{dibyac@physics.iitm.ac.in}
\emailAdd{mishaalhai97@gmail.com}
\emailAdd{tasnu94@gmail.com}
\emailAdd{ahmedrakinkamaltunok@gmail.com}
\emailAdd{sjshuvo70@gmail.com}

\abstract{In this paper, we have discovered a new avenue of fibre inflation in perturbative large volume scenario (pLVS) due to the redefinition of the base modulus. pLVS offers a novel regime where large volume of the internal space is guaranteed without the need of non-perturbative effects. In this setup, we study the possibility where a base redefinition allows to assess different versions of fibre inflation whose spectral index aligns with Atacama Cosmology Telescope (ACT) data \cite{ACT:2025fju, ACT:2025tim} and produces tensor-to-scalar ratio in the range $0.008\lesssim r\lesssim 0.01$ in different setups we have considered. The leading order flat direction - which in our case is the fibre modulus - is lifted with the combinations of string loop corrections, leading order $\alpha^{\prime  3}$$R^4$-correction, higher derivative $F^{4}$ corrections as well as our new ingredient redefinition of the modulus. Since recent Dark Energy Spectroscopic Instrument (DESI) results appear to favour a dynamical explanation for late-time acceleration over a simple cosmological constant, exploring quintessence offers a more suitable approach. In this lore, we also examine the quintessence sector to complete our model and account for both early- and late-time cosmic acceleration.  In this framework, the poly-instanton correction generates a potential along the axionic directions, and we find that the resulting quintessence behaviour and the subsequent cosmological predictions about dark matter closely resemble the predictions of the original fibre inflation scenario studied earlier.}

\begin{document} 
\maketitle
\flushbottom

\section{Introduction} 
Connecting string theory to observations is one of the main goals of string phenomenology. At the same time, our current understanding of the Universe is based on six-parameters $\Lambda$CDM model--- constructing standard model of cosmology. Our Universe provides us with the ultimate testing ground allowing us to probe various energy scales. The framework of string theory is quite robust in accommodating theories with varying energy scale requirements, ranging from very high to very low. Thus, bridging these two worlds of string theory and cosmology becomes essential for phenomenological purposes. $\Lambda$CDM has several shortcomings regarding the horizon and flatness problem (see \cite{Baumann:2009ds} for further reference), observed anisotropies in the Cosmological Microwave Background (CMB) spectrum \cite{Planck:2018jri}, and to provide the seeds for the structure formation we see today, a phenomenological addition is required. Inflation is a paradigm which can give an answer to all these issues. Our Universe, as it stands now, is going through another period of expansion, as we find galaxies moving away from each other --- dark energy provides an explanation for this expansion. Current experiments (DESI \cite{DESI:2024mwx, DESI:2024uvr, DESI:2025zgx}) suggests us to consider a dynamical dark energy like quintessence at late times.  There is a large difference in the energy scales at which these two epochs are predicted to occur.\par

Inflation in its minimal form, is formulated to be driven by a scalar field known as the inflaton that is minimally coupled to gravity. Inflation occurs as the field slowly rolls over its potential, and stops when it nears the minimum of the potential. During inflation, the energy scale can sit within a couple of orders of magnitude below the Planck scale, where Planck-suppressed operators can no longer be ignored. If primordial gravitational waves are large enough to be detectable, the inflaton must traverse a large distances in field space, which pushes any low-energy description beyond its natural radius of convergence unless the ultraviolet (UV) completion dictates the pattern of higher-dimensional corrections. In small-field models, the inflaton traverses a sub-Planckian patch, so observable tensor modes are negligible, and slow-roll then hinges on maintaining a flat potential. Exactly as is the case for inflation, dark energy too, is UV-sensitive: Planck-suppressed operators tend to steepen the potential (the $\eta$ problem), unless protected by a symmetry. String theory, being a UV-complete framework, has the ability to accommodate the necessary ingredients to explain the dynamical nature of gravity along with its quantum aspects. String theory supplies precisely the ingredients which make the study of inflation more robust.\par
Similarly, for dark energy, the cosmological constant problem is highly sensitive to UV physics because every heavy particle that gets integrated out shifts the vacuum energy by an amount set by its mass, achieveing the minuscule vacuum energy density of today impossible with purely infrared physics alone. If late-time acceleration is dynamical rather than a strict constant term, then one needs an extraordinarily light scalar, i.e. Quintessence with a mass of the order of the Hubble scale today or less, and couplings that evade fifth-force bounds. Such a task is excruciatingly difficult, requiring tools that go well beyond those able to devise the standard model, tools that can be provided by string theory in a fairly reasonable manner. In short, both inflation and dark-energy physics hinge on Planck-scale sensitivities, symmetries, and non-perturbative dynamics that only an UV-complete theory, like string theory, can specify and control.

Compactifying a ten-dimensional superstring theory on a six-dimensional compact manifold—most prominently a Calabi–Yau threefold (CY$_{3}$)—yields a four-dimensional theory with $\mathcal N=2$ supersymmetry and a large set of scalar (moduli) fields \cite{Candelas:1985en, Sen:1986pw, Nemeschansky:1986yx}. For a realistic phenomenological theory, one typically breaks to $\mathcal N=1$ by performing an orientifold projection of the CY, leading to a 4D $\mathcal N=1$ supergravity \cite{Grimm:2004uq, Grimm:2004ua}. One further requires that these scalars to generate a potential for realistic phenomenology (see for e.g.: \cite{McAllister:2023vgy, Cicoli:2023opf}). In $\mathcal N=1$ supergravity, the F-term scalar potential is determined by the Kähler potential $K$ and the holomorphic superpotential $W$. In type IIB CY orientifolds, at tree-level the superpotential (the flux superpotential) depends on the complex-structure moduli and the axio-dilaton, while the Kähler moduli do not appear at tree level; their dependence arises from non-perturbative effects such as Euclidean D3-brane instantons or gaugino condensation. By contrast, the Kähler potential receives both perturbative (in $\alpha^{\prime}$ and $g_{s}$) and non-perturbative corrections. Consequently, the tree-level dynamics fix the complex-structure moduli and the dilaton, whereas the Kähler moduli remain flat, at the level of the F-term scalar potential, at tree level due to the no-scale structure \cite{Giddings:2001yu}. The K\"ahler moduli are then lifted only once aforementioned corrections are included. 

Famously, in \cite{Kachru:2003aw}, non-perturbative effects added to the superpotential were considered, and the K\"ahler moduli were stabilised at an AdS minima. This was uplifted to dS in the KKLT scenario with the aid of an anti-D3 brane. However, this required tuning of the tree-level superpotential to be extremely small in order to compete with the sub-leading non-perturbative effect. Even though currently these small values have been engineered in \cite{McAllister:2024lnt, AbdusSalam:2025twp}, the moduli stabilisation and dS scenario can be made with a more reasonable tree-level superpotential like the one found in the the LVS scenario \cite{Balasubramanian:2005zx}. In LVS, one uses the leading $\alpha'$ correction to the K\"ahler potential to act as a \emph{balance} along with the non-perturbative corrections to the superpotential to generate an AdS minima. This is then further uplifted to a dS minima via either anti-branes \cite{Kachru:2003aw}, T-branes \cite{Cicoli:2015ylx} or D-terms \cite{Burgess:2003ic, Braun:2015pza}. 

However, one can go further in this story-telling of moduli stabilisation using perturbative corrections. Using the tree-level and one-loop correction at $\alpha^{\prime 3}$, one can stabilise the K\"ahler moduli \cite{Antoniadis:2018hqy, Antoniadis:2019rkh, Leontaris:2022rzj, Leontaris:2023obe}. This setup also gives an AdS minima, which can be uplifted via say for example D-terms. A further development was seen in \cite{Cicoli:2024bwq}, where they stabilised the K\"ahler modulus directly to a dS minimum using solely perturbative corrections courtesy of moduli redefinitions. Ref. \cite{Hai:2025wvs} then extended the analysis to multiple K\"ahler moduli. Interestingly, the tree-level superpotentential, in this perturbative stabilisation scenarios remain unconstrained due to the moduli stabilisation scheme. This brings forth a great sense of confidence when extending to inflationary scenarios. The moduli redefinitions, arising from D7-branes at gauge thresholds \cite{Conlon:2009kt,Conlon:2009qa, Conlon:2009xf}, offer an interesting possibility in forms of uplifting to dS \cite{Cicoli:2024bwq} and also in this paper where it pushes fibre inflation to ACT regime. 

With moduli stablised, one should then resort to the case of studying both the inflationary regime first. Various inflationary constructions have been proposed in the context of KKLT and LVS (see eg. \cite{Baumann:2014nda, Cicoli:2023opf} for reviews on these inflationary models). In the perturbative moduli stabilisation scheme, various inflationary scenarios have been developed over the years \cite{Bera:2024zsk, Bera:2024ihl, Bera:2024ihl, Chakraborty:2025yms, Leontaris:2025hly, Leontaris:2025xit, Cicoli:2024bwq, Hai:2025wvs}. One of the prominent inflationary scenarios is both LVS and pLVS is the Fibre Inflationary scenario\footnote{See \cite{Cicoli:2008gp, Cicoli:2016xae, Burgess:2016owb, Cicoli:2023njy, Cicoli:2024bxw} for fibre inflation in LVS context.}. However, as remarked in \cite{Ferreira:2025lrd}, generic fibre inflation models seem to predict $0.96<n_s<0.967$, but according to the current bounds set by ACT \cite{ACT:2025tim, ACT:2025fju}, the required bounds for $n_{s}$ need to be higher \footnote{A plethora of literatures have tried to address match the ACT data for example in \cite{Kallosh:2025rni, Liu:2025qca, Yogesh:2025wak, Peng:2025bws, Yin:2025rrs, Byrnes:2025kit, Wolf:2025ecy, Aoki:2025wld, Gao:2025viy, Zahoor:2025nuq, Ferreira:2025lrd, Mohammadi:2025gbu, Odintsov:2025wai, Q:2025ycf, Zhu:2025twm, Kouniatalis:2025orn, Hai:2025wvs, Dioguardi:2025vci, Yuennan:2025kde, Kallosh:2025gmg, Kallosh:2025ijd, GonzalezQuaglia:2025qem, DOnofrio:2025bol, Oikonomou:2025jmy, Odintsov:2025mqq, Odintsov:2025eiv, Oikonomou:2025xms, Odintsov:2025jky, Oikonomou:2025icu, Odintsov:2025oiy, Nojiri:2025low, Odintsov:2025jfq, Ellis:2025bzi, Yuennan:2025inm, Yuennan:2025tyx, Oikonomou:2025htz, Ellis:2025ieh, Ishiguro:2025tvn, Leontaris:2025hly, Cagan:2025rbc, Maity:2025czp}. }. In this paper, we try to remedy this situation and push fibre inflation to a more stable ACT-territory. 

We also built on the recent experimental development on understanding the late time acceleration \cite{DESI:2024uvr, DESI:2024mwx, DESI:2025zgx}. The DESI collaboration point towards a dynamical dark energy, like quintessence \cite{Ozer:1985ws, Wetterich:1987fm,Ratra:1987rm, Frieman:1995pm, Copeland:2006wr, Tsujikawa:2013fta, Andriot:2024jsh, Andriot:2024sif, Bhattacharya:2024kxp, Olguin-Trejo:2018zun, Brinkmann:2022oxy, Cicoli:2021fsd, Cicoli:2021skd, Cicoli:2024yqh, Cicoli:2020noz, Calderon-Infante:2022nxb, Alestas:2024gxe, Farakos:2020jbx}, rather than a strict cosmological constant. In light of string theory analysis, it seems that is as difficult to get quintessence (if not more) as it is to get a de-Sitter (dS) vacuum \cite{Cicoli:2018kdo}. However, a positive story can still be told as in \cite{Cicoli:2024yqh}. We consider, for our case the axion hilltop quintessence, where the axion corresponds to the fibre modulus (which is also our inflaton field). This completes our story-telling from the early universe to the current expansion phase of the universe and also accounting for some fraction of the dark matter. 

\paragraph{Summary of results and organization of the paper:}
In this paper, we consider moduli redefinition \cite{Conlon:2009qa, Conlon:2009kt, Conlon:2010ji, Klaewer:2020lfg, Weissenbacher:2020cyf} of the base which leads to three distinct inflationary scenarios. We then consider poly-instanton corrections to stabilise the corresponding axions and we observe that the axion corresponding the fibred moduli can lead to a successful quintessence scenario. The organisation of the paper is as follows:
\begin{itemize}
    \item [$\triangle$] In section~\ref{sec:2}, we first discuss the perturbative and non-perturbative corrections arising in type IIB superstring theory followed by a discussion on LVS \cite{Balasubramanian:2005zx} and pLVS \cite{Antoniadis:2018hqy, Antoniadis:2019rkh}. 
    \item [$\triangle$] section~\ref{sec:3} starts off by discussing previous fibre inflation models followed by a discussion on our new kind of fibre inflation model which has excellent agreement with recent data. 
    \item [$\triangle$] section~\ref{sec:4} discusses the stabilisation of the axions and its power to explain the dark energy. We connect this with the inflationary analysis of section~\ref{sec:3} thus providing a complete story of models capable of hosting inflation, quintessence and a portion of dark matter together. 
    \item [$\triangle$] We conclude in section~\ref{sec:5} with the discussion of our results and some open directions that we intend to pursue in the future. 
    
\end{itemize}

\section{Large Volume Scenario} \label{sec:2}

The study of moduli stabilisation in type-IIB string compactification is largely facilitated in the regime of large volume of the internal space and weak coupling, enabling us to control the perturbative corrections in string theory. In this section, we first enlist the possible perturbative corrections in a CY, discuss the standard LVS and later the perturbative LVS setup.

\subsubsection*{Perturbative corrections:}

String theory is subjected to two forms of perturbative corrections to the classical supergravity action. They are higher-derivative  $\alpha^{\prime}$ corrections\footnote{For discussions on computations of higher derivative corrections and subsequent discussions, check \cite{Burgess:2020qsc, Wulff:2024ips, Wulff:2024mgu, Wulff:2021fhr, Hsia:2024kpi, Liu:2022bfg, Aggarwal:2025lxf, Ameri:2025bei, Garousi:2024rzh}.} $\mathcal{O}(\alpha^{\prime})$ and string loop $(g_{s})$ corrections. There is no known correction to occur at $\mathcal{O}(\alpha')$. However, there do exist known corrections that occur at $\mathcal{O}(\alpha^{\prime 2})$. These are
\begin{itemize}
    \item Ref. \cite{Grimm:2013gma} computed corrections that redefine the modulus by a topological constant, leaving the K\"ahler potential unchanged. 
    \item On a similar note, at one-loop expansion in $\alpha^{\prime 2}$ i.e $\mathcal{O}\, (\alpha^{\prime \  2} g_{s})$, it has been found in \cite{Conlon:2009qa, Conlon:2009kt, Conlon:2010ji, Klaewer:2020lfg, Weissenbacher:2020cyf} that a redefinition of a modulus occur (K\"ahler modulus to be precise), and they happen to be proportional to the logarithm of the volume of the CY i.e: 
    \begin{equation}
    \tau  \rightarrow \tau'=\tau-\alpha \ln{\mathcal{V}}.
    \end{equation}
    Here, $\vol$ denotes the volume of the CY (the explicit form of which to be explained later) and $\alpha$ represents the one-loop $\beta-$function coefficient of gauge theories living on the D7 branes i.e. $$\displaystyle \alpha\sim \frac{f(\beta)}{(2\pi)^n},\,\,\,n\in \mathbb{I}$$. 
     \item Ref. \cite{Berg:2005ja, Berg:2007wt} computed exact corrections at $\mathcal{O}(\alpha^{\prime 2} g_{s}^{2})$ in case of toroidal orientifolds. However, there is a magical cancellation as seen in \cite{Cicoli:2007xp} which extends the classical no-scale structure to an extended no-scale structure. The presence of this extended no-scale structure is what makes LVS \cite{Balasubramanian:2005zx} robust. These are the KK-corrections which have an important effect, in terms of inflation, at $\mathcal{O}(\alpha'{}^4 g_{s}^4)$. 
\end{itemize}
At $\mathcal{O}(\alpha^{\prime  3})$, there are fewer corrections known. In type IIB string theory, supersymmetry actually forbids any corrections at order $\mathcal{O}(\alpha'{})$ and $\mathcal{O}(\alpha^{\prime 2})$. The first correction is actually the tree-level $\mathcal{O}(\alpha^{\prime 3})$ corrections which is proportional to the $\zeta(3)$. There are the following known corrections at $\mathcal{O}(\alpha^{\prime 3})$, whose phenomenological consequence in $4D$ supergravity\footnote{Check Appendix of \cite{Hai:2025wvs} to see a sketch of the explicit derivation.} has been extensively studied:
\begin{itemize}
    \item The tree-level $\mathcal{O}(\alpha^{\prime  3})$ $R^{4}$-term upon dimensional reduction in 4D produces a correction term which is proportional to the Euler characteristics of the CY, see \cite{Becker:2002nn} for reference. This changes the K\"ahler potential by shifting the volume by
    \begin{equation}
    \mathcal{V}\rightarrow \mathcal{V}+\frac{\xi}{2g_{s}^{3 / 2}}, \quad \xi=-\frac{\zeta(3) \chi}{2(2 \pi)^3}
    \end{equation}
    This correction is historically known as the BBHL correction which was made to compete with non-perturbative corrections in LVS to generate a minima for the volume modulus. 
    \item The log-loop correction arises at the level $\mathcal{O}(g_{s}^{2} \alpha^{\prime 3})$ and similar to the BBHL corrections, it shifts the overall volume of the CY to:
    \begin{equation}
        \mathcal{V}\to \mathcal{V}+g_{s}^{1/2}\eta\ln\mathcal{V},\,\,\eta=\frac{\chi(\text{CY})\zeta[2]}{2(2\pi)^3}
    \end{equation}

    \item There are curvature corrections at $\mathcal{O}(\alpha^{\prime 3})$ order arising from the $F^{4}$ term in 10 dimensions \cite{Ciupke:2015msa,Grimm:2017okk} which affect the F-term scalar potential as 
    \begin{equation}
    V_{F^4}=-\frac{\lambda}{64\pi^2} g_{s}^{1 / 2} \frac{\left|W_0\right|^4}{\mathcal{V}^4} \Pi_\alpha t^\alpha.
    \end{equation}
    Here, $\Pi_\alpha$ are the second Chern numbers of the CY${}_3$ defined as
    \begin{equation}\label{chern numbers}
    \Pi_i=\int_{D_i} c_2(X). 
    \end{equation}
    $\lambda$ is a small number whose value usually is $\mathcal{O}\left(10^{-2}\right)-\mathcal{O}\left(10^{-4}\right)$. 
\end{itemize}
Lastly, there are $\mathcal{O}\left(g_{s}^2 \alpha^{\prime 4}\right)$ corrections which are ruled out due to no-scale cancellation at $\mathcal{O}(\alpha^{\prime 2} g_{s}^{2})$ which arise in the case of compactification on toroidal orientifolds \cite{Berg:2005ja,Berg:2007wt}. As discussed in references \cite{Gao:2022uop, vonGersdorff:2005bf}, these corrections can also be interpreted as exchange of KK-modes between D7 (or O3-planes) and D3- branes (or O3 planes) localised in the compact dimensions. The exchange of winding strings between intersecting branes also induces a loop effect at this order. These correct the K\"ahler potential as \begin{equation}
\begin{aligned}
& V_{g_{s}}^{K K}=\frac{g_{s}^3}{8\pi\times 16} \frac{\left|W_0\right|^2}{ \mathcal{V}^4} \sum_{\alpha \beta} C_\alpha^{K K} C_\beta^{K K}\left(2 t^\alpha t^\beta-4 \mathcal{V} k^{\alpha \beta}\right) \\
& V_{g_{s}}^w=-\frac{g_{s}\left|W_0\right|^2}{4\pi\mathcal{V}^3} \sum_\alpha \frac{C_\alpha^W}{t^\alpha_{\cap} }. 
\end{aligned}
\end{equation}
Here, $k^{\alpha \beta}=\frac{\partial t^\alpha}{\partial \tau_\beta}$ and $\tau_\beta=\partial_{t^\beta} \mathcal{V}$ the 4-cycle volume.
\subsubsection*{Non-perturbative corrections:}
Non-perturbative (NP) corrections to the type IIB superpotential arise from Euclidean D3-brane (E3) instantons \cite{Witten:1996bn, Grimm:2011dj, Alexandrov:2021dyl, Alexandrov:2022mmy, Alexandrov:2021shf} and from gaugino condensation on stacks of D7-branes \cite{Gorlich:2004qm, Haack:2006cy, Kachru:2003aw}. In a 4D $\mathcal{N}=1$ compactification, one typically writes a sum of such effects
\begin{equation}
W_{\mathrm{np}}=\sum_j A_j e^{-a_j T_j},
\end{equation}
with $T_j=\tau_j+i \theta_j$ the Kähler moduli measuring the volumes of the wrapped four-cycles and their axionic partners. For an E3 instanton, one has $a_j=2 \pi$; for gaugino condensation in an $\operatorname{SU}(N)$ gauge theory on D7-branes one finds $a_j=2 \pi / N$. These terms are standard ingredients in KKLT/LVS moduli-stabilisation scenarios. More interestingly, Poly-instantons are non-perturbative effects in which the action of one instanton is corrected by another instanton \cite{Blumenhagen:2008ji, Blumenhagen:2012kz, Cicoli:2011ct}. In $\mathcal{N}=1$ type IIB orientifolds, let instanton $a$ (an E3 or gaugino condensate on D7s) generate $W_a=A_a e^{-S_a}$, while instanton $b$ corrects the holomorphic action $S_a$.
\begin{equation}
\begin{aligned}
W & =\exp \left(-S_a+S_a^{1-\text { loop }}+A_b e^{-S_b}\right) \\
& =A_a \exp \left(-S_a\right)+A_a A_b \exp \left(-S_a-S_b\right)+\ldots
\end{aligned}
\end{equation}
A standard realization (for example in \cite{Cicoli:2011ct, Cicoli:2021skd, Cicoli:2024yqh}) uses a leading NP effect on a large four-cycle $T_2$ (from D7 gaugino condensation or an E3), whose action is further corrected by an instanton on another cycle $T_1$. The resulting superpotential takes the schematic form
\begin{align}
W_{\mathrm{np}}&=A_b e^{-a_b T_b}\left(1+A_f e^{-a_f T_f}+\frac{1}{2} A_f^2 e^{-2 a_f T_f}+\cdots\right)\\
&\simeq A_b e^{-a_b T_b}+A_b A_f e^{-\left(a_b T_b+a_f T_f\right)} .
\end{align}

\subsection{The Standard LVS Setup}

The main underlying idea behind the KKLT or LVS type K\"ahler moduli stabilisation is the interplay of quantum corrections of both perturbative and non-perturbative types, utilised to stabilise the leading order flat direction. The K\"ahler direction remains flat at leading order due to the "no-scale structure" even when three-form $(F_3/H_3)$ background fluxes are turned on. The LVS setup was first posited in \cite{Balasubramanian:2005zx}, where the stabilisation was achieved via $\alpha^{\prime  3}$ corrections to the K\"ahler potential and non-perturbative corrections to the superpotential. However, non-perturbative corrections are generically quite restrictive \cite{Witten:1996bn} and depend strictly on the underlying CY geometry. It would thus be advantageous considering schemes where one fixes all K\"ahler moduli's with the aid of solely perturbative effects.\par
Fibre inflation builds on the standard LVS mechanism where the interplay of non-perturbative and leading order BBHL corrections are used to stabilise the overall volume $(\mathcal{V})$ and the blow-up divisor $(\tau_{s})$ of a fibred CY of the form:
\begin{equation}
    \mathcal{V}=\tilde{\alpha}\left(\sqrt{\tau_{f}}\tau_{b}-\gamma\tau_{s}^{3/2}\right).
\end{equation}
Here $(\tilde{\alpha},\gamma)$ are the order one constants to be fixed by the intersection numbers of the four-cycles. $\tau_{f}$ is the volume of a $T^{4}$ or $K3$ fibred over a $\mathbb{P}^1$ base. $\tau_{b}$ is the geometric modulus corresponding to the volume of the base of the fibred CY. The size of $\tau_{s}$ determines the strength of the non-perturbative effects added to the superpotential. In the limit of $\sqrt{\tau_{f}}\tau_{b} >>\tau_{s}>1$, with the help of BBHL and non-perturbative effects, the scalar potential without the uplift contribution is derived as \cite{Balasubramanian:2005zx,Cicoli:2007xp}:
\be
V_{LVS}= \frac{g_{s}}{8\pi}\frac{8a_{s}^{2}A_{s}^{2}\sqrt{\tau_{s}}e^{-2a_{s}\tau_{s}}}{3\vol}+\frac{4g_{s}|W_{0}|a_{s}A_{s}\cos(a_{s}\theta_{s})}{8\pi}\frac{\tau_{s}e^{-a_{s}\tau_{s}}}{\vol^{2}}+\frac{3\xi|W_{0}|^{2}}{32\pi\sqrt{g_{s}} \ \vol^3}.
\ee
This potential minimizes $\tau_{s}$ and a combination $\vol\sim\sqrt{\tau_{f}}\tau_{b}$: 
\be
\langle\tau_{s}\rangle\sim \xi^{2/3},\qquad \langle\vol\rangle\sim \frac{W_0\sqrt{\langle\tau_{s}\rangle}}{a_sA_s}e^{a_s\langle\tau_{s}\rangle},\qquad \langle\theta_{s}\rangle=\frac{\pi}{a_s}(1+2k_s),\,\,k_s\in \mathbb{Z}.
\ee
At this point, one needs to stabilise $\tau_{f},\; \tau_{b}$, or a combination of them that is orthogonal to $\vol$. This can be done by adding either loop corrections of either winding- or KK-type, higher derivative corrections, or log-loop corrections. This leads to the fibre inflationary scenario as discussed in the Sec. \ref{sec:3}. 

\subsection{Perturbative LVS}

In pLVS, one can choose the fibred CY $\mathbb{C} P_{[1,1,2,2,6]}^4[12]$ without the need of a blow-up divisor like LVS setups. The fundamental difference between LVS and perturbative LVS is that the latter ignores non-perturbative effects altogether due to their sub-leading nature and moduli stabilisation is done solely using perturbative corrections. The interplay here is of the tree-level $\alpha^{\prime 3}$ $R^{4}$ correction and the $\alpha^{\prime 3}$ one loop log-loop effects:
\be
V_{pLVS}= \frac{3g_{s}^{-1/2}\xi}{32\pi \vol^{3}}|W_{0}|^{2}+\frac{3g_{s}^{3/2}\eta}{16\pi\vol^{3}}\ln\vol\,|W_{0}|^{2}.
\ee
This subsequently results in an exponentially large volume, that equals to:
\be
\langle\vol\rangle\sim e^{\frac{a}{g_{s}^2}+b},\qquad a=\frac{\zeta[3]}{2\zeta[2]}\simeq 0.37,\qquad b=\frac{7}{3}.
\ee
The potential is subjected to an uplift to ensure a Minkowski minimum. Many such mechanisms exist to do so, with one being via the utilisation of anti-branes \cite{Kachru:2003aw, Balasubramanian:2005zx}, D-terms \cite{Burgess:2003ic, Braun:2015pza} and T-branes \cite{Cicoli:2015ylx} with a contribution 
\begin{equation}
V_{u p}=\frac{\delta_{u p}}{\mathcal{V}^{n / 3}}, \quad n<9, \quad \delta_{u p}>0 .
\end{equation}

\section{Fibre Inflation} \label{sec:3}
\subsection{Original Fibre Inflation}
Fibre inflation arises in the context of type IIB Calabi-Yau compactifications where the internal geometry admits a K3 fibration (or $T^4$) over a $\mathbb{P}^{1}$ base \cite{Cicoli:2008gp}. In the LARGE Volume Scenario (LVS), complex-structure moduli and the axio-dilaton are fixed by fluxes, while K\"ahler moduli are stabilised by the interplay of the $\alpha^{\prime 3}$ correction to the Kähler potential and non-perturbative superpotential terms \cite{Balasubramanian:2005zx} as discussed in the preceeding section. For fibred CYs at hand, this leaves the fibre modulus flat at leading order (extended no-scale cancellations \cite{Cicoli:2007xp}), so its potential is generated by string-loop effects; which is precisely what the original fibre-inflation construction exploits.\par
As summarised in \cite{Chakraborty:2025yms}, there are four possible scenarios available in the literature which are able to lift the flat directions. The original FI~\cite{Cicoli:2008gp}, utilises string loop correction of winding and KK type. The second type of FI \cite{Cicoli:2016chb} uses winding loops and $F^4$ correction. The third type of FI \cite{Cicoli:2016xae} utilises loop corrections (both KK and winding type) and $F^4-$terms. Finally, the fourth type \cite{Bera:2024ihl} which has direct relevance to the present work, which uses log-loop correction to stabilise the overall CY volume and utilises winding loops and higher derivative corrections to stabilise the leading order flat direction. In the next subsection, we discuss the perturbative large volume stabilisation mechanism in details which is at the heart of fibre inflation studied in the fourth type \cite{Bera:2024ihl}. Thus, including all these corrections, one obtains an inflationary potential for the fibred modulus which goes as follows
\begin{equation}
\begin{aligned}\label{fibre_OG}
V_{\mathrm{inf}} & =V_{g_{s}}^{K K}+V_{g_{s}}^W+V_{F^4} \\
& =\frac{g_{s}^3}{8\pi\times 16} \frac{\left|W_0\right|^2}{ \mathcal{V}^4} \sum_{\alpha \beta} C_\alpha^{K K} C_\beta^{K K}\left(2 t^\alpha t^\beta-4 \mathcal{V} k^{\alpha \beta}\right)-\frac{g_{s}\left|W_0\right|^2}{4\pi\mathcal{V}^3} \sum_\alpha \frac{C_\alpha^W}{t^\alpha_{\cap} }-\frac{\lambda}{64\pi^2} \sqrt{g_s} \frac{\left|W_0\right|^4}{\mathcal{V}^4} \Pi_\alpha t^\alpha \\
& = \frac{\left|W_0\right|^2}{\mathcal{V}^2}\Bigg(\frac{g_{s}^3}{32\pi} \frac{\left(C_f^{K K}\right)^2}{\tau_f^2}+ \frac{g_s^3}{8\pi\times 72}\frac{\left( C_b^{K K}\right)^2 \tau_f}{\mathcal{V}^2}- \frac{ g_{s} C^W}{8\pi\mathcal{V} \sqrt{\tau_f}}    \\
& \hspace{1.4em} +\frac{3}{8\pi^2}\sqrt{g_s}W_0^2C^{F^4}_1 \frac{1}{\tau_f \mathcal{V} }+ \frac{9}{16\pi^2}\sqrt{g_s}W_0^2C^{F^4}_2 \frac{\sqrt{\tau_f}}{ \mathcal{V}^2} 
\Bigg) .
\end{aligned}
\end{equation}

\subsection{The Effect of Moduli Redefinition in Fibre Inflation}

As discussed in Sec. \ref{sec:2}, moduli redefinitions can occur at 1-loop and the corresponding modulus gets redefined as
 \begin{equation}
    \tau  \rightarrow \tau'=\tau-\alpha \ln{\mathcal{V}}.
    \end{equation}
In case of a CY with $h^{1,1}=1$ \cite{Cicoli:2024bwq}, it was found that a redefinition of the modulus can act as a natural perturbative uplift to a dS which enabled them to solve the infamous $\eta$-problem of brane-antibrane inflation \cite{Burgess:2001fx, Kachru:2003sx, Villa:2025zmj}. This provided a natural mechanism to obtain a dS vacua while using solely perturbative corrections. This was also generalised to $h^{1,1}>1$ in \cite{Hai:2025wvs} for swiss-cheese CYs. In case of a fibred CYs, one can redefine both the base and the fibre. A redefintion of the base will lead to a K\"ahler potential:
\begin{equation}\label{redef-base}
  K=- 2 \ln{\sqrt{\tau_f}(\tau_b - \alpha  \ln{\sqrt{\tau_f}\tau_b})}
\end{equation}
where, $\alpha$ is the moduli redefinition parameter. In the following sub-section, we shall see how a redefinition of the base leads to different versions of fibre inflation with great concordance with ACT data \cite{ACT:2025fju,ACT:2025tim,Kallosh:2025ijd}.
For completeness, a redefinition of the fibre will change the K\"ahler potential as:
\begin{equation}
    K=- 2 \ln{\sqrt{\left(\tau_f- \alpha  \ln{\sqrt{\tau_f}\tau_b}\right)}\tau_b },
\end{equation}
This produces a leading term in the F-term scalar potential $V(\tau_f) \propto \frac{\alpha}{\tau_f^2 \tau_b^2}=\frac{\alpha}{\tau_f \mathcal{V}^2}$. For example, this is in principle a stronger effect compared to the string loop correction $V(\tau_f) \propto g_s^2 \frac{(c_f^{KK})^2}{\tau_f^2 \mathcal{V}^2}$ which has a further $g_s^2$ suppression alongside a $\tau_f$ and the square of a small parameter $c_f^{KK}$. This would require an extremely small value of moduli redefinition parameter $\alpha$ in order for it not to compete with the leading $\alpha^{\prime}$ needed to stabilise the volume.

\subsection{New Inflationary Scenarios}
As emphasized earlier, in this subsection we examine how redefining the base modulus of a $K3-$fibred CY can give rise to a new class of fibre inflation models. Our model is able to match not only the Planck \cite{Planck:2018jri} but also the more recent ACT data \cite{ACT:2025fju, ACT:2025tim}. With this effect included, we investigate the resulting fibre inflation scenarios that can emerge and yield interesting inflationary predictions. We shall show that each fibre inflation type studied in this paper differ  from the original fibre inflation model \cite{Cicoli:2008gp} with the key distinctions along the inflationary observables.  The authors of \cite{Cicoli:2018cgu,Cicoli:2020bao} have also obtained a higher values of $n_s$ closer to current ACT data (which was not available back then) without the base-modulus redefinition. 

\subsection*{Case 1} To lift the leading order flat direction \ie the fibre modulus in this case, the scenario at hand features loop-corrections of KK and winding type as well as our main distinctive correction known as moduli redefinition of the base. In other words, it incorporates the perturbative corrections same as ~\cite{Cicoli:2008gp}, but includes base redefinition. The following situations may arise when the corresponding topological numbers of the $\mathcal{O}(F^4)$-correction \eqref{chern numbers} are zero or they are very small and thus negligible. The potential thus takes the form:
\begin{equation} \label{scn1pot}
    V(\tau_f) = \frac{\left|W_0\right|^2}{\mathcal{V}^2}\left(\frac{g_{s}^3}{32\pi} \frac{\left(c_f^{K K}\right)^2}{\tau_f^2}-\frac{g_{s}}{8\pi} \frac{ c^W}{\mathcal{V} \sqrt{\tau_f}}+\frac{3g_s\alpha \sqrt{\tau_f}}{8\pi\mathcal{V}}\right).
\end{equation}

The above potential \eqref{scn1pot} has its volume stabilised by the leading order log-loop corrections. Now, we want to stabilise $\tau_{f}$. We denote the vacuum expectation value (vev) of the fibre modulus to be $\langle\tau_f\rangle=\tau_{f_{*}}$. Using the canonical normalisation $\tau_f=\tau_{f_{*}} e^{k\phi}$ where we have $k=\frac{2}{\sqrt{3}}$ and $\phi$ denoting the inflaton field, we can choose an uplift term $\delta_{up}$ to obtain a Minkowski vacuum for the potential \eqref{scn1pot}. To achieve this, we require the potential and the first derivative of the potential to be zero  at $\phi=0$. Applying canonical normalization and adding an uplift term changes the potential of \eqref{scn1pot} to\footnote{Note that the moduli redefinition parameter $\alpha$ of \cite{Cicoli:2024bwq} and $\alpha$ in this article differ as $\alpha_{\text{present}}=\frac{g_{s}}{8\pi} \alpha_{\text{before}}$.}:
\begin{equation} \label{scn1up}
     V(\phi) = \frac{\left|W_0\right|^2}{\mathcal{V}^2}\left(\frac{g_{s}^3}{32\pi} \frac{\left(C_f^{K K}\right)^2e^{-2k\phi}}{\tau_{f_{*}}^2}-\frac{g_{s}}{8\pi} \frac{ C^W e^{-k\phi/2}}{\mathcal{V} \sqrt{\tau_{f_{*}}}}+\frac{3g_s\alpha}{8\pi}\frac{ \sqrt{\tau_{f_{*}}} e^{k\phi/2}}{\mathcal{V}}\right)+\delta_{up}.
\end{equation}
For notational convenience, we write 
\begin{equation}
    a=\frac{\left|W_0\right|^2g_{s}^3 \left(C^{KK}_f\right)^2}{32\pi\mathcal{V}^2\tau_{f_{*}}^2},\quad b=\frac{3\alpha \left|W_0\right|^2g_s\sqrt{\tau_{f_{*}}}}{8\pi\mathcal{V}^3}, \quad d=\frac{\left|W_0\right|^2 g_{s} C^W}{8\pi\mathcal{V}^3\sqrt{\tau_{f_{*}}}},
\end{equation}
Imposing $V(\phi)|_{\phi=0}=0$ gives us: 
\begin{equation}
    \delta_{up}=d-a-b
\end{equation}
and requiring $V^{\prime}(\phi)|_{\phi=0}=0$ leads to 
\begin{equation}
 a=\frac{1}{4}(b+d).
\end{equation}
Combining these two conditions, one can write $\delta_{up}$ as:
\begin{equation}
\delta_{up} =\frac{3d}{4}-\frac{5b}{4}
\end{equation}
Plugging this form of $\delta_{up}$ back in \eqref{scn1up} gives the final inflationary potential under the first case as where note that $\tau_{f_{\star}}$ has been put to its numerically calculated vev:
\begin{equation} \label{sit1upp}
  V_{\mathrm{inf}}(\phi)
  = b \left(e^{\frac{k\phi}{2}} + \frac{1}{4} e^{-2k\phi} - \frac{5}{4}\right)
  + d\left(\frac{3}{4} - e^{-\frac{k\phi}{2}} + \frac{1}{4} e^{-2k\phi}\right)
\end{equation}
We demonstrate an illustrative plot for this potential below for different values of the parameter $\alpha$ emphasizing the strength of moduli redefinition. 
\begin{figure}[H]
    \hspace{1.5cm}\includegraphics[width=1.1\linewidth]{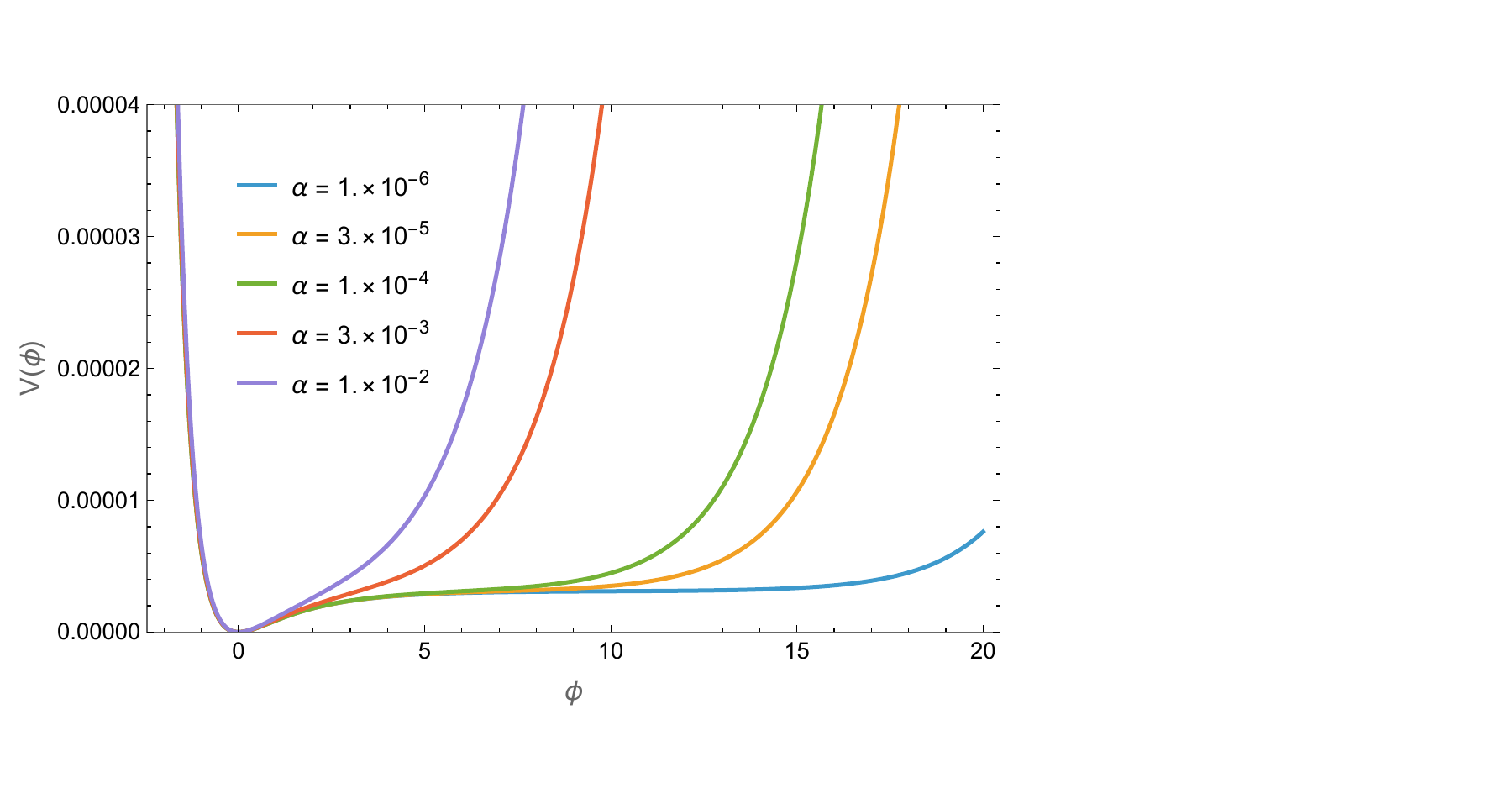}
    \caption{Plot of potential $V(\phi)$ in \eqref{sit1upp} for varying values of $\alpha$ with minima of $\tau_{f}$ is at $1.27$. For this plot, we set $g_{s}=0.4, \  c_{f}^{KK}=0.04, \ c^{w}=0.2, \ |W_{0}|=12.85$. The minima of $\tau_{f}$ remain essentially unchanged for different values of $\alpha$. }
    \label{fig:pot1}
\end{figure}
We notice that different parametric values for $\alpha$ help flatten the tail-end of the potential. The smaller the value of $\alpha$, the flatter the tail-end becomes as is evident in Fig.~\ref{fig:pot1}. Interestingly, the minimum of the potential is independent of the value of $\alpha$. As the potential \eqref{sit1upp} features a flat plateau, it will be useful in generating a viable model of inflation as we review now. 

To begin our inflationary analysis, let us introduce some of the basic definitions. The slow-roll parameters of inflation are defined through Hubble parameter as well as the potential as:
\begin{align}
   & \epsilon_H=-\frac{\dot{H}}{H^2}=\frac{1}{H}\frac{dH}{dN},\quad \eta_H=\frac{\dot{\epsilon_H}}{\epsilon_H H}=\frac{1}{\epsilon_H}\frac{d\epsilon_H}{dN} \\
   & \epsilon_V=\frac{1}{2}\left(\frac{V^{\prime}}{V}\right)^2\quad\quad \eta_V=\left(\frac{V^{\prime\prime}}{V}\right)\label{slow-roll-pot}
\end{align}
where the dot denotes time derivative and $N$ denotes the number of e-folds determined by,
\begin{equation}
N(\phi)=\int H\,dt=\int_{\phi_{end}}^{\phi_{\star}}\frac{1}{\sqrt{2\epsilon_H}}d\phi\simeq \int_{\phi_{end}}^{\phi_{\star}}\frac{V}{V'}d\phi,
\end{equation}
where $\phi_{\star}$ is the point of horizon exit of the pivot scale at which the cosmological observables are currently matched with the experimental data. $\phi_{end}$ on the other hand denotes the end of inflation when the value of slow-roll parameters become one. For single field models of inflation, such as the one studied in this paper, the slow-roll parameters are related as:
\be
\epsilon_V=\epsilon_H\left(1+\frac{\eta_H}{2(3-\epsilon_H)}\right)^2\simeq \epsilon_H,\quad \eta_H\simeq -2\eta_V+4\epsilon_V
\ee
The cosmological observables such as the amplitude of the scalar power spectrum, the spectral index, as well as tensor-to-scalar ratio can be computed in terms of the potential slow-roll parameters defined in \eqref{slow-roll-pot} as :
\begin{align}\label{inf_obs_bound}
&P_{s} \equiv \frac{V_{inf}^{\star}}{24\pi^{2}\epsilon_{H}^{\star}}\simeq 2.2\times 10^{-9}, \quad \text{or}\quad \frac{V_{inf}^{\star \ 3}}{V_{inf}^{\prime 2}} \simeq 2.6\times 10^{-7},\nn\\
&n_{s}-1 = -2\epsilon_{H}^{\star}-\eta_{H}^{\star} \simeq 2 \eta_{V}^{\star}-6 \epsilon_{V}^{\star} \simeq -0.04,\nn\\
&r = 16\epsilon_{H}^{\star} \simeq 16 \epsilon_{V}^{\star}.
\end{align}
The bound on $n_s$ from the Planck data \cite{Planck:2018jri} is $n_s=0.9651\pm 0.0044$ and the upper bound on the tensor-to-scalar ratio according to the joint analysis of Planck and BICEP/Keck \cite{BICEP:2021xfz} $r$ is $r<0.036$. However, more recently, the ACT data along with the Planck data and the BAO data from DESI observation have slightly shifted the spectral index towards a bluer tilt with $n_s=0.9743\pm 0.0034$.\par
In order to formulate a successful period of inflation, we numerically solve the background and Friedmann equations:
\begin{align}
&\Ddot{\phi}+3H\dot{\phi}+V_{,\phi}=0,\\
& H^2=\frac{1}{3M_{pl}^2}\left(\frac{1}{2}\dot{\phi}^2+V(\phi)\right)
\end{align}
The slow-roll parameters in case 1 take the following form 
\begin{equation}
    \begin{split}
        \epsilon&=\frac{e^{-\frac{8 \phi }{\sqrt{3}}} \left(b \left(e^{\frac{5 \phi }{\sqrt{3}}}-1\right)+d \left(e^{\sqrt{3} \phi }-1\right)\right)^2}{6 \left(b \left(\frac{1}{4} e^{-\frac{4 \phi }{\sqrt{3}}}+e^{\frac{\phi }{\sqrt{3}}}-\frac{5}{4}\right)+\frac{1}{4} d \left(e^{-\frac{4 \phi }{\sqrt{3}}}-4 e^{-\frac{\phi }{\sqrt{3}}}+3\right)\right)^2}\\
        \eta&=\frac{4 b \left(e^{\frac{5 \phi }{\sqrt{3}}}+4\right)-4 d \left(e^{\sqrt{3} \phi }-4\right)}{3 \left(-5 b e^{\frac{4 \phi }{\sqrt{3}}}+4 b e^{\frac{5 \phi }{\sqrt{3}}}+b-4 d e^{\sqrt{3} \phi }+3 d e^{\frac{4 \phi }{\sqrt{3}}}+d\right)}.
    \end{split}
\end{equation}
Although $\{\epsilon,\eta\}$ do not depend on the overall $W_0$ factor, it does depend on the constant coefficients appearing in front of the perturbative corrections such as $\{c_f^{KK},c_W\}$.
We make a plot of the slow-roll parameters below, which suggests the flatness of the potential over a large range of $\phi$. 
\begin{figure}[H]
    \centering
    \includegraphics[width=0.75\linewidth]{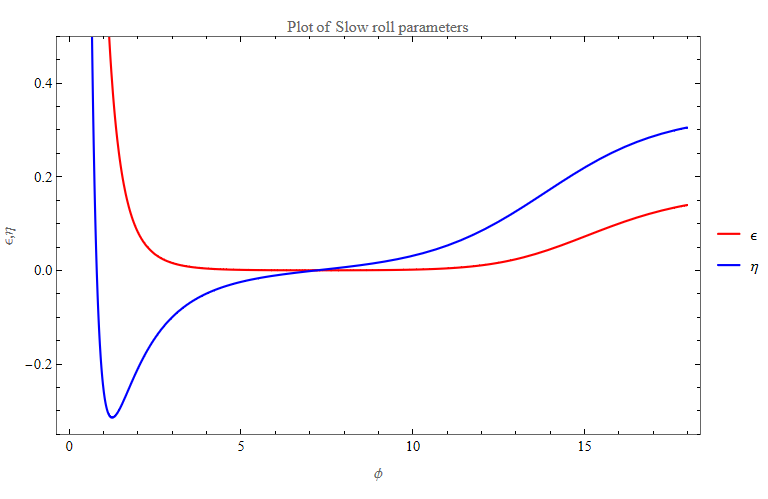}
    \caption{Slow-roll parameters ($\epsilon$ and $\eta$) vs. $\phi$. For comparative purposes, the same parametric values were considered as in Fig. \ref{fig:pot1} with $\alpha\sim10^{-5}$. This plot illustrates that both parameters become small over the field range relevant for slow-roll inflation.}
    \label{fig:2}
\end{figure}
We now present a set of parameters to illustrate the range of values that can yield viable inflationary scenarios along with observable predictions
\begin{table}[H]
\begin{center}
\centering
    \resizebox{0.9\textwidth}{!}{ 
    \begin{tabular}{| l | c | c | c | c | c | c | c | c | c | c | c | c | c | }
\hline
\cellcolor[gray]{0.9} Sub-case &\cellcolor[gray]{0.9} $|W_0|$ &  \cellcolor[gray]{0.9} $g_{s}$ &  \cellcolor[gray]{0.9} $\xi$ &  \cellcolor[gray]{0.9} $C_w$ &  \cellcolor[gray]{0.9} $\alpha$ &  \cellcolor[gray]{0.9} $C_f^{KK}$ &  \cellcolor[gray]{0.9} $\langle\tau_f^{\star}\rangle$ &  \cellcolor[gray]{0.9} $\langle\vol\rangle$ \\
\hline \hline
 $1$ & $12.85$ & $0.4$ & $1.5$ & $0.2$ & $1.11\times 10^{-5}$ & $0.04$ & $1.27$ & $1119.72$    \\
\hline
 $2$ & $45$ & $0.42$ & $1.8$ & $0.1$ & $0.33\times 10^{-6}$ & $0.07$ & $5.98$ & $1694.89$   \\
\hline
 $3$ & $230$ & $0.4$ & $2$ & $0.2$ & $0.33\times 10^{-5}$ & $0.071$ & $7.74$ & $5341.88$    \\
\hline
\end{tabular}}
\end{center} 
\caption {Benchmark parameters for the potential \eqref{sit1upp}.}
\label{tab1}
\end{table}
 In all these sub-cases, parameters are varied to make sure in each case the correct value for the amplitude of the scalar curvature power spectrum is matched with the observational predictions of \eqref{inf_obs_bound}. The vev of $\tau_{f}$ is obtained such that it falls within the range selected by its dark energy counterpart as we show in Sec. \ref{sec:4}.   In all the subcases, we have maintained the hierarchy between the leading $\alpha'$ corrections and the other sub-leading corrections. This is needed while lifting the leading-order flat direction \ie the fibre direction at a fixed volume. This very fact gives a certain bound on the coefficient of the sub-leading corrections as given below:
\begin{align}\label{main_bound}
    & C^W\ll\frac{3\xi \sqrt{\tau_{f\star}}}{4g_s^{3/2}},\quad (C_f^{KK})^2\ll\frac{3\xi \tau_{f\star}^2}{g_s^{7/2}\mathcal{V}},\quad (C_b^{KK})^2\ll\frac{54\mathcal{V}\xi}{g_s^{7/2}\tau_{f\star}},\nonumber\\
    & C_1^{F^4}\ll\frac{\pi \xi \tau_{f\star}}{4 g_s W_0^2},\quad C_2^{F^4}\ll\frac{\pi\xi\mathcal{V}}{6g_s W_0^2 \sqrt{\tau_{f\star}}},\quad \alpha\ll\frac{\xi}{4g_s^{3/2}\sqrt{\tau_{f\star}}}.
\end{align}

For the parameters presented in table \ref{tab1}, we make sure that the bounds from \eqref{main_bound} on the coefficients are satisfied.\footnote{For example, for values in \ref{tab1}, we get from \eqref{main_bound} that $\alpha<1.9$ for sub-case 1. However, the actual values needed for a successful inflationary scenario is $\alpha\sim \mathcal{O}(10^{-4})$ which is trivially satisfied.}We emphasize the effect of moduli redefinition of the base to have a distinct effect on the inflationary observables as we show below in Fig.~\ref{fig:placeholder1}. As evident in Fig.~\ref{fig:placeholder1}, the model prediction deviates from the Planck contour and enters in the $1\sigma$ domain of the ACT data. Across all cases, including the potential in \eqref{potmain}, we observe that the overall factor $W_0^2$ remains unfixed in the calculation of slow-roll parameters and hence it allows for a more straightforward matching of the scalar power-spectrum amplitude.

\begin{figure}[H]
    \centering
    \includegraphics[width=0.75\linewidth]{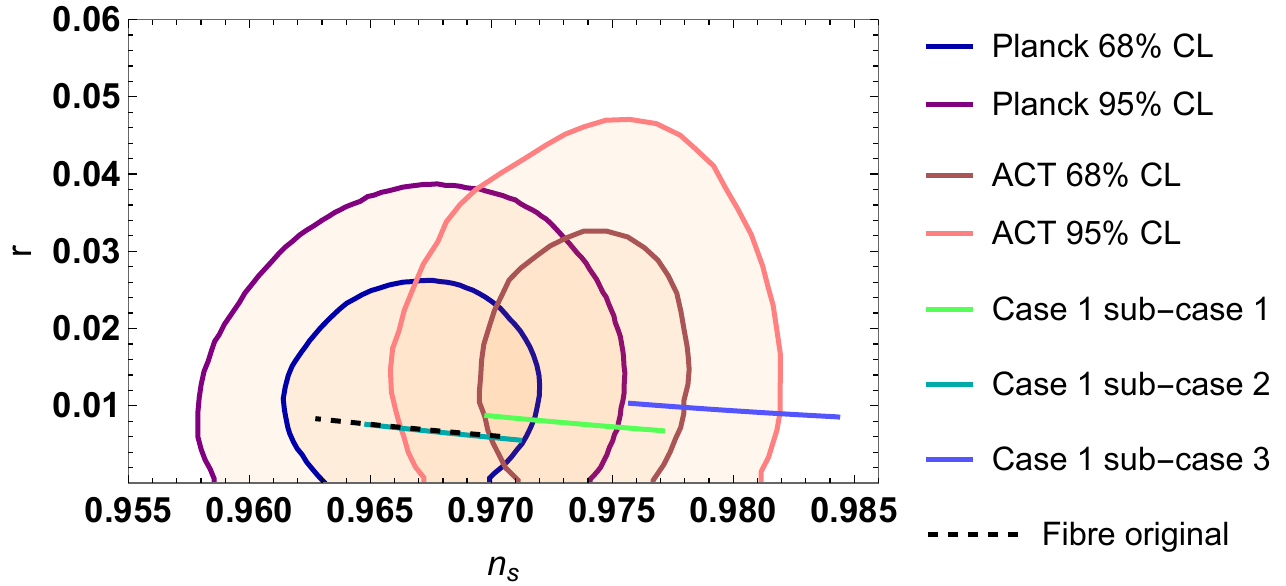}
    \caption{The above visualises the trend for $r$ vs. $n_{s}$, computed for various sub-cases at 50-60 e-foldings. The dotted line denoting the original fibre model \cite{Cicoli:2008gp} and the colourful solid lines denoting sub-cases from Table \ref{tab1}.}
    \label{fig:3}
\end{figure}
As is clear from Fig.~\ref{fig:3}, this model explores a wider range in the $(n_s,r)-$plane. Most importantly, it aligns more towards the ACT data still being in the weak coupling and large volume limit. 
\subsection*{Case 2} 
In this case, we analyze the presence of winding type loop correction and $F^4$ corrections along with redefinition of the base modulus. This scenario arises because loop corrections of KK type generally enjoy an extended no-scale structure as shown in ~\cite{Cicoli:2007xp}. 
In a scenario where the KK-loop corrections are absent, the potential takes the following form:
\begin{equation}
    V(\tau_f) = \frac{\left|W_0\right|^2}{\mathcal{V}^2}\left(\frac{3g_s\alpha}{8\pi}\frac{ \sqrt{\tau_f}}{\mathcal{V}}-\frac{g_{s}}{16\pi} \frac{2 C^W}{\mathcal{V} \sqrt{\tau_f}}+\frac{3}{8\pi^2}\sqrt{g_s}W_0^2C^{F^4}_1 \frac{1}{\tau_f \mathcal{V}} + \frac{9}{16\pi^2}\sqrt{g_s}W_0^2C^{F^4}_2 \frac{\sqrt{\tau_f}}{ \mathcal{V}^2 }\right).\label{sit2pot}
\end{equation}
For convenience we can rewrite the above potential as:  
\begin{equation}
    a=\frac{3g_s\alpha |W_0|^2\sqrt{\tau_{f_{*}}}}{8\pi\mathcal{V}^3}, \quad b=\frac{ g_{s} |W_0|^2 C^W}{8\pi\mathcal{V}^3 \sqrt{\tau_{f_{*}}}},\quad c=\frac{3\sqrt{g_s}|W_0|^4 C^{F^4}_1}{8\pi^2\mathcal{V}^3\tau_{f_{*}}}, \quad d=\frac{9\sqrt{g_s}|W_0|^4 C^{F^4}_2 \sqrt{\tau_{f_{*}}}}{16\pi^2\mathcal{V}^4}.
\end{equation}
The volume is minimised using the leading order pLVS mechanism and the vev of $\tau_f$ is found numerically. Following the workflow of the previous section the uplift term is found as 
\begin{equation}
    \delta_{up}=c-2(a+d).
\end{equation}
So the inflationary potential in this case becomes 
\begin{equation}
  V_{\mathrm{inf}}(\phi)=\left(a+d\right)\left(e^{\frac{k \phi}{2}}+e^{\frac{-k \phi}{2}}-2\right)+c\left(e^{-k \phi}-2 e^{\frac{-k\phi}{2}}+1\right)\label{sit2potup}
\end{equation}
We give an illustrative plot of this potential below: 
\begin{figure}[H]
    \hspace{1.5cm}
    \includegraphics[width=1.1\linewidth]{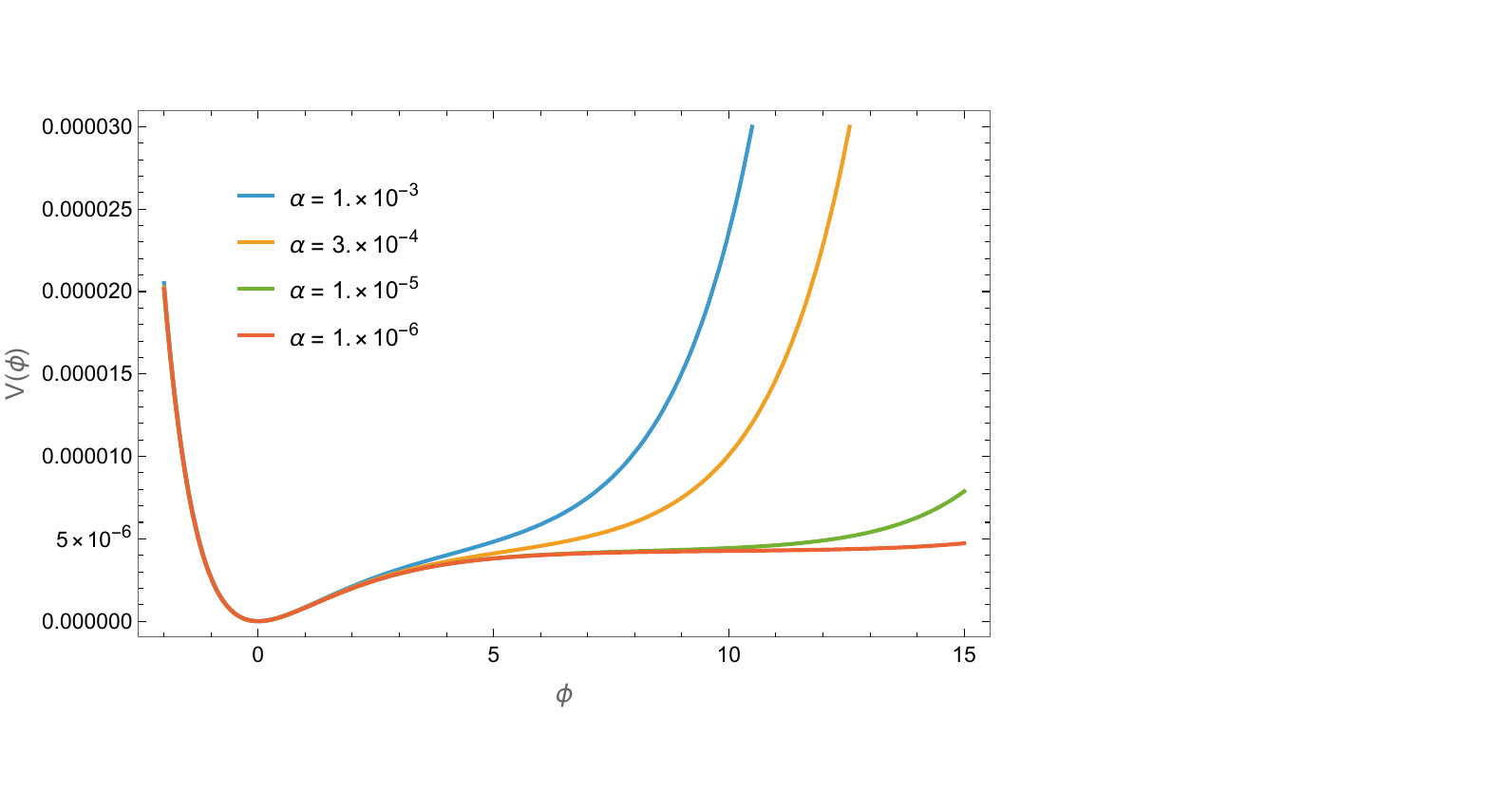}
    \caption{Illustrative plots showing case 2 potential for different values of $\alpha$. Increasing $\alpha$ mainly steepens the large $\phi$ tail.}
    \label{fig:placeholder1}
\end{figure}
Slow-roll parameters then take the subsequent form 
\begin{equation}
\begin{split}
\epsilon&=\frac{\left(-e^{\sqrt{3} \phi } (a+d)-2 c \left(e^{\frac{\phi }{\sqrt{3}}}-1\right)+e^{\frac{\phi }{\sqrt{3}}}\right)^2}{6 \left(e^{\sqrt{3} \phi } (a+d)+c \left(e^{\frac{\phi }{\sqrt{3}}}-1\right)^2+e^{\frac{\phi }{\sqrt{3}}}-2 e^{\frac{2 \phi }{\sqrt{3}}}\right)^2},\\
\eta&=\frac{e^{\sqrt{3} \phi } (a+d)-2 c \left(e^{\frac{\phi }{\sqrt{3}}}-2\right)+e^{\frac{\phi }{\sqrt{3}}}}{3 \left(e^{\sqrt{3} \phi } (a+d)+c \left(e^{\frac{\phi }{\sqrt{3}}}-1\right)^2+e^{\frac{\phi }{\sqrt{3}}}-2 e^{\frac{2 \phi }{\sqrt{3}}}\right)},
\end{split}
\end{equation}
which again depends on coefficients of the quantum corrections through $\{a,c,d\}$.
\begin{table}[H]
\begin{center}
\centering
    \resizebox{0.9\textwidth}{!}{ 
    \begin{tabular}{| l | c | c | c | c | c | c | c | c | c | c | c | c | c | }
\hline
\cellcolor[gray]{0.9} Sub-case &\cellcolor[gray]{0.9} $W_0$ &  \cellcolor[gray]{0.9} $g_{s}$ &  \cellcolor[gray]{0.9} $\xi$ &  \cellcolor[gray]{0.9} $C_w$ &  \cellcolor[gray]{0.9} $\alpha$ &  \cellcolor[gray]{0.9} $C_1^{F^4}$ & \cellcolor[gray]{0.9}$C_2^{F^4}$ &  \cellcolor[gray]{0.9} $\langle\tau_f^{\star}\rangle$ &  \cellcolor[gray]{0.9} $\langle\vol\rangle$   \\
\hline \hline
 $1$ & $0.99$ & $0.392$ & $0.34$ & $0.607$ & $0.33\times 10^{-5}$ & $0.438$ & $10^{-5}$ & $4.65$ & $212.601$  \\
\hline
 $2$ & $0.9$ & $0.408$ & $0.97$ & $0.6$ & $0.33\times 10^{-5}$ & $0.54$ & $10^{-5}$  & $4.75$ & $189.972$  \\
\hline
 $3$ & $1$ & $0.4$ & $1$ & $0.9$ & $0.33\times 10^{-5}$ & $0.68$ & $10^{-5}$  & $5.2$ & $234.706$  \\
\hline
\end{tabular}}
\end{center} 
\caption {Benchmark parameters for the potential \eqref{sit2potup}.}
\label{tab2}
\end{table}

\begin{figure}[H]
    \centering
    \includegraphics[width=0.75\linewidth]{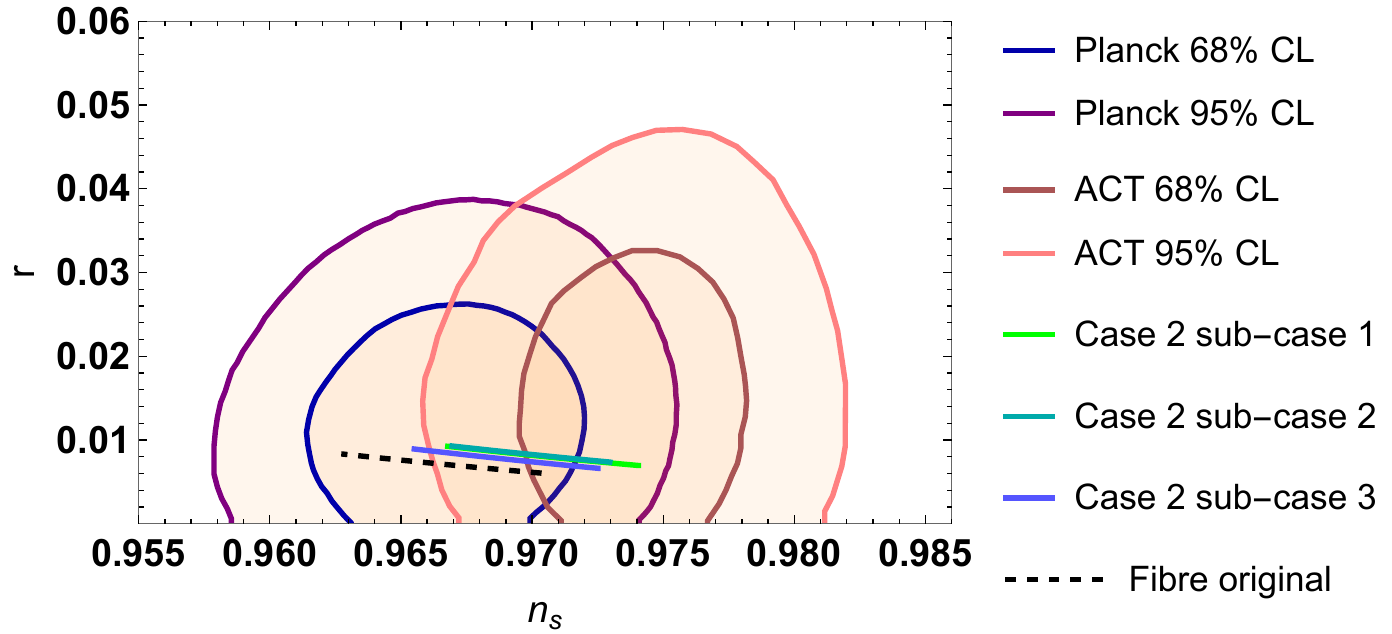}
    \caption{$r$ vs. $n_{s}$ plot in similar fashion to Fig. \ref{fig:3}. Dotted lines again denote the original fibre model \cite{Cicoli:2008gp}, with the solid lines denoting sub-cases from Table \ref{tab2}.}
    \label{fig:slowrollsit1}
\end{figure}

Similar to case $1$, this model also deviates away from the original fibre inflation --- pushing it towards the ACT contours. The interplay of the loop corrections, higher derivative curvature corrections and moduli redefinition helps us to move the predictions away from the Planck data towards the ACT data. 
\subsection*{Case 3}

\begin{figure}[H]
    \hspace{1cm}
    \includegraphics[width=0.75\linewidth]{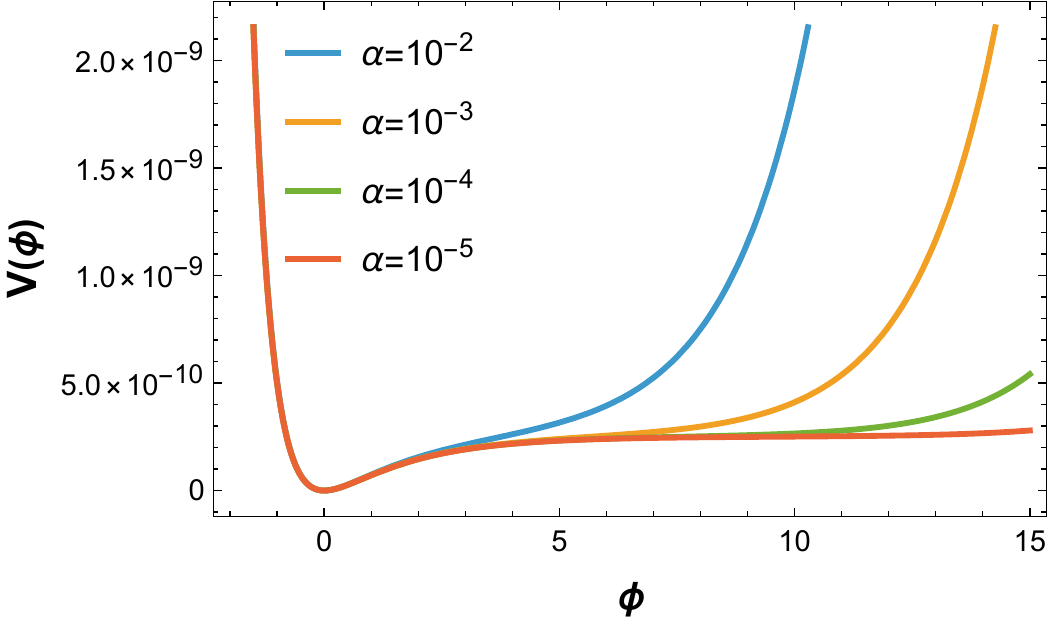}
    \caption{Inflationary potential for case 3.}
    \label{fig:pot3}
\end{figure}

In this setup, alongside moduli redefinitions, we consider the winding correction and KK-correction corresponding to the fibred modulus and one of the $\mathcal{O}(F^4)$ terms. The potential becomes:
\begin{equation}\label{pot4}
    V(\tau_f) = \frac{\left|W_0\right|^2}{\mathcal{V}^2}\left(\frac{3g_s\alpha}{8\pi}\frac{ \sqrt{\tau_f}}{\mathcal{V}}-\frac{g_{s}}{16\pi} \frac{2 C^W}{\mathcal{V} \sqrt{\tau_f}}+\frac{3}{8\pi^2}\sqrt{g_s}W_0^2C^{F^4}_1 \frac{1}{\tau_f \mathcal{V} }+\frac{g_{s}^3 \left(C^{KK}_f\right)^2}{32\pi\tau_f^2}\right) .
\end{equation}
Following the similar path as before, after stabilizing $\tau_f$ we adjust the uplift term given by
\begin{equation}
    \delta_{up}=\frac{3b}{4}-\frac{c}{2}-\frac{5a}{4} 
\end{equation}
where,
\begin{equation}
    a=\frac{3\alpha g_s|W_0|^2 \sqrt{8\pi\tau_{f_{*}}}}{\mathcal{V}^3}\quad b=\frac{2 g_{s} C^W |W_0|^2}{16\pi\mathcal{V}^3\sqrt{\tau_{f_{*}}}}\quad c=\frac{2\sqrt{g_s}C^{F^4}_{1}|W_0|^4}{8\pi^2\mathcal{V}^3\tau_{f_{*}}}\quad d=\frac{g_{s}^3 |W_0|^2\left(C^{KK}_{f}\right)^2} {32\pi\mathcal{V}^2\tau^2_{f_{*}}}
\end{equation}
The inflationary potential after including the uplift, becomes 
\begin{equation}\label{pot3}
    V(\phi)=a\left(\frac{1}{4} e^{-2 k \phi }+e^{\frac{k \phi }{2}}-\frac{5}{4}\right)+b \left(\frac{1}{4} e^{-2 k \phi }-e^{-\frac{1}{2} (k \phi )}+\frac{3}{4}\right)+c \left(-\frac{1}{2} e^{-2 k \phi }+e^{-k \phi }-\frac{1}{2}\right)
\end{equation}
The slow roll parameters for this potential are
\begin{equation}
\begin{split}
    \epsilon&=\frac{\left(a \left(\frac{e^{\frac{\phi }{\sqrt{3}}}}{\sqrt{3}}-\frac{e^{-\frac{4 \phi }{\sqrt{3}}}}{\sqrt{3}}\right)+b \left(\frac{e^{-\frac{\phi }{\sqrt{3}}}}{\sqrt{3}}-\frac{e^{-\frac{4 \phi }{\sqrt{3}}}}{\sqrt{3}}\right)+c \left(\frac{2 e^{-\frac{4 \phi }{\sqrt{3}}}}{\sqrt{3}}-\frac{2 e^{-\frac{2 \phi }{\sqrt{3}}}}{\sqrt{3}}\right)\right)^2}{2 \left(a \left(\frac{1}{4} e^{-\frac{4 \phi }{\sqrt{3}}}+e^{\frac{\phi }{\sqrt{3}}}-\frac{5}{4}\right)+b \left(\frac{1}{4} e^{-\frac{4 \phi }{\sqrt{3}}}-e^{-\frac{\phi }{\sqrt{3}}}+\frac{3}{4}\right)+c \left(-\frac{1}{2} e^{-\frac{4 \phi }{\sqrt{3}}}+e^{-\frac{2 \phi }{\sqrt{3}}}-\frac{1}{2}\right)\right)^2}\\
    \eta&=\frac{a \left(\frac{4}{3} e^{-\frac{4 \phi }{\sqrt{3}}}+\frac{e^{\frac{\phi }{\sqrt{3}}}}{3}\right)+b \left(\frac{4}{3} e^{-\frac{4 \phi }{\sqrt{3}}}-\frac{1}{3} e^{-\frac{\phi }{\sqrt{3}}}\right)+c \left(\frac{1}{3} (-8) e^{-\frac{4 \phi }{\sqrt{3}}}+\frac{4}{3} e^{-\frac{2 \phi }{\sqrt{3}}}\right)}{a \left(\frac{1}{4} e^{-\frac{4 \phi }{\sqrt{3}}}+e^{\frac{\phi }{\sqrt{3}}}-\frac{5}{4}\right)+b \left(\frac{1}{4} e^{-\frac{4 \phi }{\sqrt{3}}}-e^{-\frac{\phi }{\sqrt{3}}}+\frac{3}{4}\right)+c \left(-\frac{1}{2} e^{-\frac{4 \phi }{\sqrt{3}}}+e^{-\frac{2 \phi }{\sqrt{3}}}-\frac{1}{2}\right)},
\end{split}
\end{equation}
which again depends on the model parameters but not on the overall $\{W_0,\vol\}$, for different values of $\alpha$, we generate plots of the above potential in Fig.~\ref{fig:pot3}. 
We choose three sub cases for the case $3$ which we show below:
\begin{table}[H]
\begin{center}
\centering
    \resizebox{0.9\textwidth}{!}{ 
    \begin{tabular}{| l | c | c | c | c | c | c | c | c | c | c| c |c |c |c |c |}
\hline
\cellcolor[gray]{0.9} Sub-cases & \cellcolor[gray]{0.9} $g_{s}$ &\cellcolor[gray]{0.9} $W_0$ &  \cellcolor[gray]{0.9} $d$ &  \cellcolor[gray]{0.9} $\xi$ &  \cellcolor[gray]{0.9} $C_w$ &  \cellcolor[gray]{0.9} $\alpha$ &  \cellcolor[gray]{0.9} $C_1^{F^4}$ &  \cellcolor[gray]{0.9} $C_2^{F^4}$ &  \cellcolor[gray]{0.9} $C_f^{KK}$ &  \cellcolor[gray]{0.9} $\langle\tau_f^{\star}\rangle$ & \cellcolor[gray]{0.9 }$\langle\vol\rangle$   \\
\hline \hline
 $1$ & $0.42$ & $7$ & $1$ & $1.6$ & $0.
 8$ & $ 0.33\times10^{-5}$ & $10^{-4}$ & $10^{-5}$ & $0.3$ & $7.12$ & $961.487$  \\
\hline
 $2$ & $0.359$ & $40$ & $1$ & $1.4$ & $0.4$ & $0.33\times 10^{-5}$ & $10^{-4}$ & $10^{-5}$ & $0.1$ & $1.8$ & $2355.95$ \\
\hline
 $3$ & $0.45$ & $9$ & $1$ & $2.05$ & $0.25$ & $0.33\times 10^{-5}$ & $10^{-4}$ & $3*10^{-5}$ & $0.08$ & $0.1$ & $1627.9$ \\
\hline
\end{tabular}}
\end{center} 
\caption {Benchmark parameters for the potential \eqref{pot3}.}
\label{tab3}
\end{table}
The inflationary prediction for the potential \eqref{pot3} is shown below: 
\begin{figure}[H]
    \centering
    \includegraphics[width=0.75\linewidth]{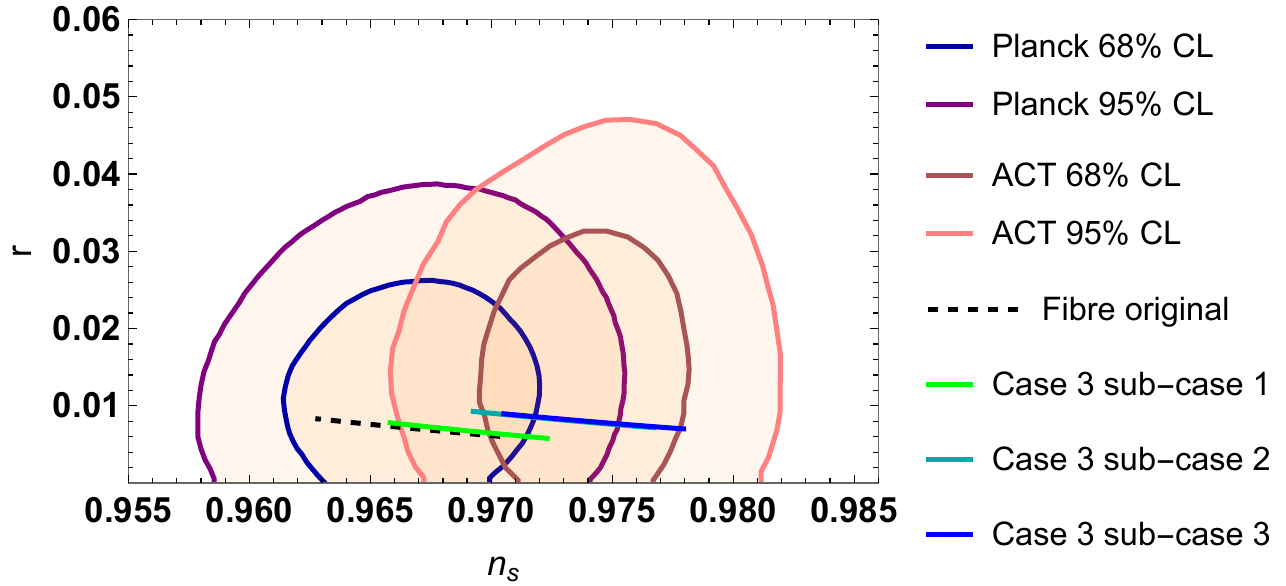}
    \caption{Constraints in the $n_s,r$ plane for case 3. The colourful solid segments correspond to the two sub-cases of scenario 3, illustrating how different parameter choices shift the model within the observationally allowed region. Again, dashed black line represents original fibre inflation \cite{Cicoli:2008gp}.}
    \label{figure_7}
\end{figure}
From Fig.~\eqref{figure_7}, we observe that one of the sub-cases differ from the original FI scenario. In all the sub-cases presented we duly satisfy the bounds of \eqref{main_bound}.
\subsection*{Case 4}
This is the final case presented in this article, incorporating all the sub-leading corrections discussed so far to lift the leading order flat direction. They are loop corrections of KK- and winding-type, higher derivative $F^4$ correction as well as moduli redefinition:
\begin{align}\label{potmain}
   & V_{\mathrm{inf}} = \frac{\left|W_0\right|^2}{\mathcal{V}^2}\Bigg(g_{s}^3 \frac{\left(C_f^{K K}\right)^2}{32\pi\tau_f^2}+\frac{3g_s\alpha \sqrt{\tau_f}}{8\pi\mathcal{V}}+\frac{g_{s}^3}{16\pi\times 72} \frac{2\left(C_b^{K K}\right)^2 \tau_f}{\mathcal{V}^2}\nonumber\\
   &\hspace{0.8cm}-g_{s} \frac{2 C^W}{16\pi\mathcal{V} \sqrt{\tau_f}}+\frac{3}{8\pi^2}\sqrt{g_s}W_0^2C^{F^4}_1 \frac{1}{\tau_f \mathcal{V} }+ \frac{9}{16\pi^2}\sqrt{g_s}W_0^2C^{F^4}_2 \frac{\sqrt{\tau_f}}{ \mathcal{V}^2 }\Bigg).
\end{align}

 For the numerical analysis, we have considered several benchmark models for the potential type \eqref{potmain}:
\begin{table}[H]
\begin{center}
\centering
    \resizebox{0.9\textwidth}{!}{ 
    \begin{tabular}{| l | c | c | c | c | c | c | c | c | c | c | c | c |c |c |c |c |}
\hline
\cellcolor[gray]{0.9} Sub-cases & \cellcolor[gray]{0.9} $g_{s}$ &\cellcolor[gray]{0.9} $W_0$ &  \cellcolor[gray]{0.9} $d$ &  \cellcolor[gray]{0.9} $\xi$ &  \cellcolor[gray]{0.9} $C_w$ &  \cellcolor[gray]{0.9} $\alpha$ &  \cellcolor[gray]{0.9} $C_1^{F^4}$ &  \cellcolor[gray]{0.9} $C_2^{F^4}$ &  \cellcolor[gray]{0.9} $C_f^{KK}$ &  \cellcolor[gray]{0.9} $C_b^{KK}$ &  \cellcolor[gray]{0.9} $\langle\tau_f^{\star}\rangle$ &  \cellcolor[gray]{0.9} $\langle\vol\rangle$    \\
\hline \hline
 $1$ & $0.35$ & $7$ & $1$ & $1.051$ & $0.3$ & $0.33 \times 10^{-5}$ & $10^{-3}$ & $10^{-4}$ & $0.0589$ & $0.002$ & $1.51$ & $752.296$   \\
 \hline
 $2$ & $0.34$ & $10$ & $1$ & $1$ & $0.3$ & $0.33\times 10^{-5}$ & $10^{-3}$ & $10^{-4}$ & $0.045$ & $0.01$ & $1.96$ & $779.453$ \\
\hline
 $3$ & $0.47$ & $5.55$ & $1$ & $1.85$ & $0.3$ & $0.33 \times 10^{-5}$ & $10^{-3}$ & $10^{-4}$ & $0.05$ & $0.08$ & $2.48$ & $679.086$\\
\hline
\end{tabular}}
\end{center} 
\caption {Benchmark parameters for the potential \eqref{potmain}.}
\label{tab4}
\end{table}

Due to the complicated form of \eqref{potmain}, we minimize the potential with respect to the fibre direction numerically. We now show an illustration:
\begin{figure}[H]
    \centering
    \includegraphics[width=0.75\linewidth]{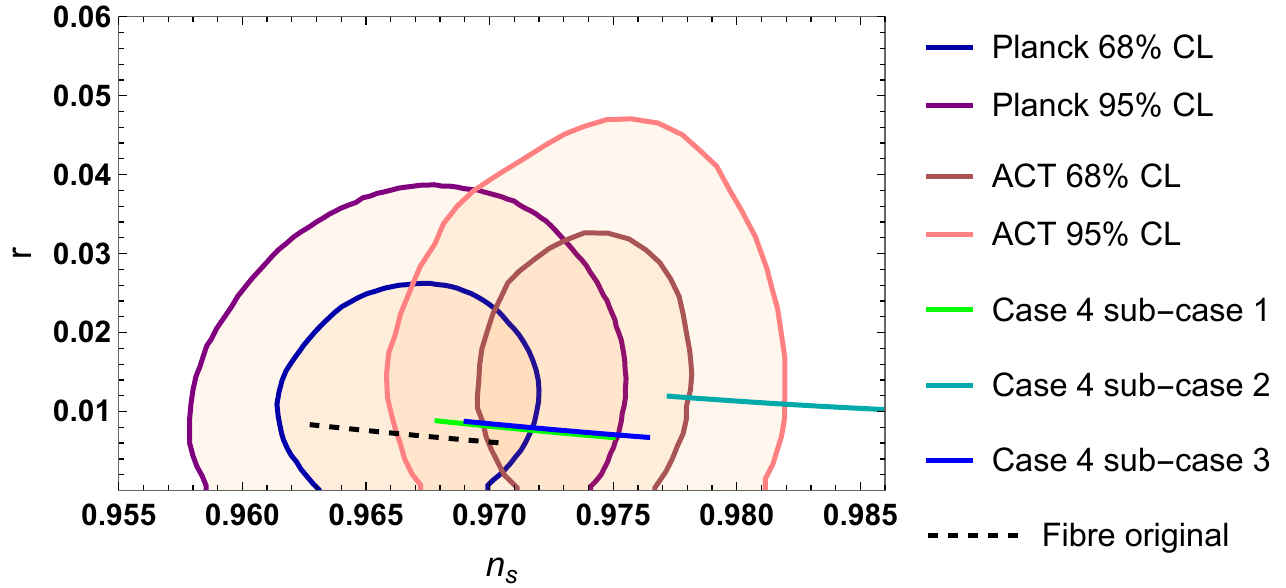}
    \caption{The behavior of spectral index versus tensor-to-scalar ratio for the fibre potential \eqref{potmain}. In this case, we notice our model prediction differs substantially with the original fibre inflation \cite{Cicoli:2008gp} prediction with an affinity towards the ACT data.}
    \label{fig:8}
\end{figure}
In this case, when all the corrections are taken into account, we clearly see that the sub-cases differ sufficiently from the original fibre inflation prediction. The deviation from the original fibre prediction denotes the fact that the base redefinition actually plays a pivotal role in the overall inflationary observables. Similar to other cases, the case 4 also satisfies the bounds on the parameters given in \eqref{main_bound}.
\section{Quintessence from Ultra-Light Axions} \label{sec:4}

Until now, the storytelling has been focused on moduli stabilisation and inflation with an emphasis on the new addition of moduli redefinition of base in the context of fibre inflation. In this lore, we follow the standard three-step procedure where the heavier modulus \ie the volume direction is stabilised with the help of leading order corrections. This is done by tree-level $\alpha^{\prime  3}$ $R^{4}$ correction and non-perturbative effects in case of LVS and tree-level $\alpha^{\prime  3}$ $R^{4}$ correction together with log-loop correction in the case of pLVS. In this article, we followed the approach of a pLVS setup to stabilise the volume direction. As a next step, we identify the leading order flat direction and lift it with the help of sub-leading corrections of higher derivative $F^{4}-$correction, loop corrections of winding and KK type and finally moduli redefinition of one of the K\"ahler modulus. In the preceding subsection, we mainly showed how various interplay between these sub-leading corrections actually push the fibre inflationary scenario towards interesting regions currently favoured by the observational data \cite{ACT:2025fju,ACT:2025tim}. DESI results \cite{DESI:2024mwx, DESI:2024uvr, DESI:2025zgx} point towards a dynamical dark energy, which is why we try to find a quintessence field in our setup which can aid us in this regard. Thus, as a final step, we introduce poly-instanton corrections (as discussed in Sec. \ref{sec:2}) to lift the flat directions along the axions of the corresponding K\"ahler moduli which are already stabilised in the previous two steps.\par
The primary motivation for examining the role of poly-instanton effects in axion stabilisation is that these axions can serve as progenitors not only for a quintessence field—providing a dynamical dark energy component—but also for contributing to a portion of dark matter abundance. Following the footsteps of \cite{Cicoli:2024yqh}, the quintessence potential then becomes: 
\begin{equation}
\begin{aligned}
V_{\text {late }}= & -\frac{g_{s}\left|W_{0}\right| A_{b}}{2\pi \mathcal{V}^2}\left(a_{b} \tau_{b}\right) e^{-a_{b} \tau_{b}} \cos \left(a_{b} \theta_{b}\right) \\
& -\frac{g_{s}\left|W_{0}\right| A_{b} A_{f}}{2\pi\mathcal{V}^2}\left(a_{b} \tau_{b}+a_{f} \tau_{f}\right) e^{-\left(a_{b} \tau_{b}+a_{f} \tau_{f}\right)} \cos \left(a_{b} \theta_{b}+a_{f} \theta_{f}\right)
\end{aligned}
\end{equation}
where we omitted the redefinition of base as in \eqref{redef-base} because the coefficient $\alpha$ appearing as an overall constant is usually small which makes the negative exponential term even smaller. A more familiar form of the above potential is:
\begin{equation}
V_{\text {late }}=\Lambda_b^{4}\left[1-\cos \left(a_{b} \theta_{b}\right)\right]+\Lambda_{f}^{4}\left[1-\cos \left(a_{b} \theta_{b}+a_{f} \theta_{f}\right)\right]
\end{equation}
where,
\begin{equation}
\Lambda_{b}^{4} \equiv \frac{g_{s}\left|W_{0}\right| A_{b}}{2\pi\mathcal{V}^{2}} a_{b}\left\langle\tau_{b}\right\rangle e^{-a_{b}\left\langle\tau_{b}\right\rangle} \gg \Lambda_{f}^{4} \equiv \Lambda_{b}^{4}\left(1+\frac{a_{f}\left\langle\tau_{f}\right\rangle}{a_{b}\left\langle\tau_{b}\right\rangle}\right) A_{f} e^{-a_{f}\left\langle\tau_{f}\right\rangle}
\end{equation}
This can be written as the leading order potential
\begin{equation}
V_{\text {late }} \simeq \Lambda_{b}^{4}\left[1-\cos \left(\frac{\phi_{b}}{f_{b}}\right)\right]
\end{equation}
where we used the canonical normalisation $\theta_{b}=\left\langle\tau_{b}\right\rangle \phi_{b}$ and $\Lambda_{b}>>\Lambda_{f}$. We also consider $A_{b}=1=A_{f}$ and $a_{i} n=2\pi$.
The decay constant for the base and fibre axion are defined through:
\begin{equation}\label{decay_const}
f_b \equiv \frac{N_b}{2 \pi\left\langle\tau_b\right\rangle}, \qquad f_f \equiv \frac{N_f}{2\sqrt{2} \pi\left\langle\tau_f\right\rangle}
\end{equation}
While the axion $\phi_f$ associated with the sub-leading potential drives quintessence, the leading potential and its axion \ie $\phi_b$ will contribute to the dark matter. The masses of the axions are given by:
\be\label{axion_masses}
m_{b}\simeq \frac{\Lambda_{b}^{2}}{f_{b}},\qquad m_{f}\simeq \frac{\Lambda_{f}^{2}}{f_{f}}.
\ee
At leading order the putative quintessence direction $\phi_{f}$ is flat and can lead to quintessence once $\phi_{b}$ sits at it's minimum $\l \phi_{b} \r =0$. The sub-leading potential is then
\begin{equation}
V\left(\phi_{f}\right) \simeq \Lambda_{f}^4\left[1-\cos \left(\frac{\phi_{f}}{f_{f}}\right)\right]
\end{equation}
Note that the heavy axion $\phi_{b}$ can oscillate around the minima and amount to dark matter abundance. The dark energy scale will be given by
\begin{equation}
\Lambda_{f}^{4} \simeq 10^{-120} M_{p}^{4}
\end{equation}
Since $\Lambda_{f}$ is generated from a poly-instanton correction, one can explain the smallness of it more efficiently. 

With this at hand, we look into our four inflationary cases and connect them to this late-time acceleration phase of our universe. Following \cite{Cicoli:2024yqh}, the first thing we need to check is the decay constants for the quintessence axion and the dark matter axion. For that, we introduce $\Delta_{\text{max}}(f_{f})$ which tells us the maximum allowed displacement of the axion from the hilltop of a potential sustaining quintessence. At the same time, incorporating the swampland constraints on the decay constant, we bound the decay constant as $f_{f} \in (0.02, 0.1)M_{P}$. The lower bound depends on $f_{f}$, which is model dependent. We will measure this parameter for each inflationary case studied in Sec. \ref{sec:3}. The authors in ~\cite{Cicoli:2021skd,Cicoli:2024yqh} explained the relation between the quintessence axion decay constant and inflationary scale in terms of $\Delta_{\text{max}}$ and it is given by:
\begin{equation}
\ln \Delta_{\max }=c_{0}+c_{1} \ln f_{a}+c_{2}\left(\ln f_{a}\right)^{2}
\end{equation}
with $c_{0}=-32.6, c_{1}=-28.977$ and $c_{2}=-8.2302$. For the quintessence field to not be displace from it's initial position, we need 
\begin{equation}
\Delta_{\max }\left(f_{f}\right)>H_{\mathrm{inf}}
\end{equation}

The heavier axion, which in our case is the base modulus, sources the dark matter in our Universe. Similar to \cite{Cicoli:2024yqh}, this axion cannot compose all the dark matter budget of the Universe. \par
We will now present the subsequent late time predictions for the four sub-cases studied in the previous sections.

\begin{table}[H]
\begin{center}
\centering
    \resizebox{0.9\textwidth}{!}{ 
    \begin{tabular}{| l | c | c | c | c | c | c | c |}
\hline
\cellcolor[gray]{0.9} Cases & \cellcolor[gray]{0.9} $H_{inf}^{\star}\,(M_{pl})$ &\cellcolor[gray]{0.9} $f_f\,(M_{pl})$ &  \cellcolor[gray]{0.9} $f_b\,(M_{pl})$ &  \cellcolor[gray]{0.9} $m_f (eV)$ &  \cellcolor[gray]{0.9} $m_b (eV)$ &  \cellcolor[gray]{0.9} $N_b$ &  \cellcolor[gray]{0.9} $N_f$   \\
\hline \hline
 $1$ & $9.86 \times 10^{-6}$ & $0.083$ & $0.004$ & $2.89 \times 10^{-32}$ & $8.64 \times 10^{-27}$ & $\simeq 16$ & $\simeq 6$   \\
\hline
 $2$ & $8.804 \times 10^{-6}$ & $0.082$ & $0.004$ & $2.93\times 10^{-32}$ & $1.19 \times 10^{-26}$ & $\simeq 11$ & $\simeq 4$   \\
\hline
 $3$ & $8.11 \times 10^{-6}$ & $0.082$ & $0.004$ & $2.96 \times 10^{-32}$ & $9.24 \times 10^{-27}$ & $\simeq 7$ & $\simeq 5$   \\
\hline
 $4$ & $8.97 \times 10^{-6}$ & $0.082$ & $0.004$ & $2.93 \times 10^{-32}$ & $9.65 \times 10^{-27}$ & $\simeq 133$ & $\simeq 1$   \\
\hline
\end{tabular}}
\end{center} 
\caption {Parameters of quintessence era corresponding to the four kinds of inflationary scenarios. }
\label{tab5}
\end{table}
In the table above, we have chosen one sub-case under each case from Sec. \ref{sec:3}. In Table \ref{tab5}, the symbol $\star$ represents Hubble exit of the pivot scale. We calculate the decay constant of the lighter axion and the heavier axion using \eqref{decay_const} and their corresponding masses through \eqref{axion_masses}. As noted in Table \ref{tab5}, the hierarchy between the decay constant $f_f>f_b$ invariably indicates that $m_b>m_f$ and base modulus is the dark matter axion and fibre is the quintessence axion. Since for all cases, mass of the dark matter axion $m_b$ remains higher than $10^{-27}\,\mathrm{eV}$, this axion will  produce slightly higher percentage of the full dark matter budget of the Universe compared to the previously studied model in \cite{Cicoli:2024yqh}. At the same time, as tabulated in Table \ref{tab5} one of the sub-cases the rank of the gauge group correspond to the dark matter axion takes a slightly larger value $\mathcal{O}(100)$ but as shown in \cite{Louis:2012nb} the actual rank may depend on the explicitness of the underlying CY geometry. A point to note that the values of $N_b$ are dependent on $\tau_b$ so if one tries to go to higher values of volume, then $N_b$ becomes larger. Lastly, since the values of $H_{\text{inf}}$ for the cases examined in Sec. ~\ref{sec:3} differ only mildly from one another---and also remain close to the $\Delta_{\max}(f_f)$ profile reported in \cite{Cicoli:2021skd}---we conclude that although our model predictions are distinct (differing by atleast an order of magnitude) from the previously studied FI setup, they remain largely consistent across the different cases considered in this work.

\section{Conclusion and Outlook}\label{sec:5}
In this paper, we extended the original fibre inflationary scenario by considering the moduli redefinition of the base of a fibred CY while stabilising the volume modulus by using only perturbative corrections to the K\"ahler potential. This leads to a novel fibre inflationary scenario which has excellent agreement with the current data \cite{ACT:2025fju, ACT:2025tim, Kallosh:2025gmg, Kallosh:2025ijd}. With the redefinition of the base as a new ingredient, we have analysed four sub-cases by a careful interplay between various sub-leading perturbative corrections. The overall volume is stabilised following pLVS mechanism where the presence of leading order $\alpha^{\prime  3}\,R^4$ corrections, together with log-loop correction at $\mathcal{O}(\alpha'{}^3 g_{s}^2)$ order generates the desired large volume of the underlying CY. In this fibred CY, the volume is stabilised by pLVS and fibre remains as a leading order flat direction.\par
In order to lift this direction, we analyse four distinct cases featuring an interplay between several perturbative corrections with moduli redefintion being a common ingredient among all the these candidates. Among the four cases studied as presented in Sec. \ref{sec:3}, case $1$ features loop corrections of KK and winding type similar to the original fibre inflation studied in ~\cite{Cicoli:2008gp}. As we have shown in Sec. \ref{sec:3}, the effect of redefinition of the base modulus is successful in having a non-trivial effect at the level of the F-term potential which helps this case $1$ to comply completely with the recent data, unlike the original fibre models studied earlier. Similarly, for the case $2$, we assess the effects of winding loop corrections and higher derivative $F^4-$corrections along with the redefinition of the modulus. For case $3$, we allow an interplay of loop corrections of both KK and winding type, higher derivative $F^4$ corrections and base redefinition. Finally, case $4$ can be considered as a summation of all of these four cases where we have loop corrections of both types, higher derivative corrections and base modulus redefinition. As our principle results, we find in all these cases, our inflationary observables is in full agreement with the current data, whereas the fibre inflationary scenario \cite{Cicoli:2008gp} has model prediction satisfying the Planck data \cite{Planck:2018jri} but not the more recent ACT data. In this sense, our model acts like an improvement of the well-known fibre inflationary model studied in the literature \cite{Cicoli:2008gp}.\par

We confirm this claim by analysing several benchmark models. For each of these sub-cases we calculate both the spectral index and the tensor-to-scalar ratio whereas keeping the correct value of the amplitude of the scalar power spectrum. In all these cases and their sub-cases, we have made sure that the scalar potential has the correct hierarchy of sub-leading effects added. We have made this possible by choosing small values for the overall coefficients associated with each sub-leading corrections. \par

We also study the quintessence scenario in these class of models and conclude that, a successful axion hilltop quintessence can occur if one considers poly-instanton corrections to the super-potential. An important point to note is that, since we are using perturbative corrections to stabilise the moduli, these instanton and poly-instantonic corrections are quite sub-leading and do not effect the inflationary and moduli stabilisation analysis of the saxions. One also notes that, the axionic partner of the base modulus can explain a small portion of the dark matter abundance. Lastly, we believe one can expand our work in the following way:

\begin{itemize}
    \item [$\triangle$] Perturbative moduli stabilisation offers a new avenue to study cosmological scenarios. In our work, we connected both inflation and quintessence together in a coherent way with a small dark matter fraction. A next step is to check the reheating story and also the rest of the dark matter abundance. Since enough dark matter cannot be generated by the axion of base modulus, one needs to check for other dark matter candidates like primordial black-holes \cite{Cicoli:2018asa} as fibre inflation is rich enough to accommodate that. 

    \item [$\triangle$] Another interesting avenue is that we only considered fibre inflation in perturbative stabilisation. However, other inflationary scenarios, which have an $\eta$-problem due to non-perturbative moduli stabilisation, like brane-antibrane inflation \cite{Burgess:2001fx, Majumdar:2003da, Cicoli:2024bwq, Villa:2025zmj, Majumdar:2002hy} and axion monodromy inflation \cite{McAllister:2008hb} ($\eta$-problem for b-axion) can also be considered to evade their $\eta$-problem through perturbative stabilisation like \cite{Cicoli:2024bwq}. Like our case of a fibred CY, one can study quintessence in these setups. These models have very rich phenomenology and can explain problems like dark matter (through both axions and PBH), cosmic strings \cite{Majumdar:2002hy, Davis:1999tk, Majumdar:2005qc, Brunelli:2025ems} and also observable gravitational waves \cite{Ghoshal:2025tlk, Villa:2025wgl}.  
\end{itemize}
We hope to return to address some of these questions in the near future. 
\section*{Acknowledgements}
ARK is supported by Czech Science Foundation GAČR grant "Dualities and higher derivatives" (GA23-06498S) and acknowledges mental support from Kozel and Radegast during dark times. ARK specifically thanks CERN theory group and Irene Valenzuela for the hospitality at CERN. MSJS acknowledges support from NSF grant 2209116. STJ and MH are supported by the Research Seed Grant Initiative 2025 (RSGI) of BRAC University. STJ would like to acknowledge her father for always inspiring her.  MH would like to thank Francisco Gil Pedro for all the stimulating discussions in String phenomenology and Abu Mohammad Khan for his support and his unflinching assistance along the way. ARK, STJ, MH and MSJS would also like to acknowledge M. Fahim Hoque for illuminating discussions.
\appendix
\bibliography{biblio}
\bibliographystyle{JHEP}
\end{document}